\documentclass[sigconf]{acmart}

\usepackage{bm}
\usepackage{mathtools}
\usepackage{multirow}
\usepackage{color}
\usepackage{enumitem}
\usepackage[utf8]{inputenc}

\def\projname{HOOV}
\def\optitrack{\textsc{OptiTrack}}
\newcommand{\oculus}{\textsc{OCULUS}}
\def\studyone{\textsc{Study~1}}
\def\studytwo{\textsc{Study~2}}
\def\hoov{\textsc{HOOV}}

\def\drop{\textsc{DROP}}
\def\grab{\textsc{GRAB}}
\def\compound{\textsc{COMPOUND}}

\def\degree{{\footnotesize$^{\circ}$}}
\DeclarePairedDelimiter\abs{\lvert}{\rvert}

\AtBeginDocument{%
  \providecommand\BibTeX{{%
    \normalfont B\kern-0.5em{\scshape i\kern-0.25em b}\kern-0.8em\TeX}}}

\copyrightyear{2023} 
\acmYear{2023} 
\setcopyright{acmlicensed}\acmConference[CHI '23]{Proceedings of the 2023 CHI Conference on Human Factors in Computing Systems}{April 23--28, 2023}{Hamburg, Germany}
\acmBooktitle{Proceedings of the 2023 CHI Conference on Human Factors in Computing Systems (CHI '23), April 23--28, 2023, Hamburg, Germany}
\acmPrice{15.00}
\acmDOI{10.1145/3544548.3581468}
\acmISBN{978-1-4503-9421-5/23/04}

\begin{document}

%
%
\title[HOOV: Hand Out-Of-View Tracking for Proprioceptive Interaction using Inertial Sensing]{HOOV: Hand Out-Of-View Tracking for Proprioceptive Interaction using Inertial Sensing}

\author{Paul Streli}
\affiliation{%
 \institution{Department of Computer Science \\ ETH Zürich
    \country{Switzerland}}
}

\author{Rayan Armani}
\affiliation{%
 \institution{Department of Computer Science \\ ETH Zürich
    \country{Switzerland}}
}

\author{Yi Fei Cheng}
\affiliation{%
 \institution{Department of Computer Science \\ ETH Zürich
    \country{Switzerland}}
}

\author{Christian Holz}
\orcid{0000-0001-9655-9519}
\affiliation{%
 \institution{Department of Computer Science \\ ETH Zürich
    \country{Switzerland}}
}


\renewcommand{\shortauthors}{Streli et al.}

\begin{abstract}
Current Virtual Reality systems are designed for interaction under visual control.
Using built-in cameras, headsets track the user's hands or hand-held controllers while they are inside the field of view.
Current systems thus ignore the user's interaction with \emph{off-screen content}---virtual objects that the user could quickly access through proprioception without requiring laborious head motions to bring them into focus.
In this paper, we present \emph{HOOV}, a wrist-worn sensing method that allows VR users to interact with objects \emph{outside} their field of view.
Based on the signals of a single wrist-worn inertial sensor, HOOV continuously estimates the user's hand position in 3-space to complement the headset's tracking as the hands leave the tracking range.
Our novel data-driven method predicts hand positions and trajectories from just the continuous estimation of hand orientation, which by itself is stable based solely on inertial observations.
Our inertial sensing simultaneously detects finger pinching to register off-screen selection events, confirms them using a haptic actuator inside our wrist device, and thus allows users to select, grab, and drop virtual content.
We compared HOOV's performance with a camera-based optical motion capture system in two folds.
In the first evaluation, participants interacted based on tracking information from the motion capture system to assess the accuracy of their proprioceptive input, whereas in the second, they interacted based on HOOV's real-time estimations.
We found that HOOV's target-agnostic estimations had a mean tracking error of 7.7\,cm, which allowed participants to reliably access virtual objects around their body without first bringing them into focus. 
We demonstrate several applications that leverage the larger input space HOOV opens up for quick proprioceptive interaction, and conclude by discussing the potential of our technique.
\end{abstract}



\begin{teaserfigure}
    \centering
    \includegraphics[width=\textwidth]{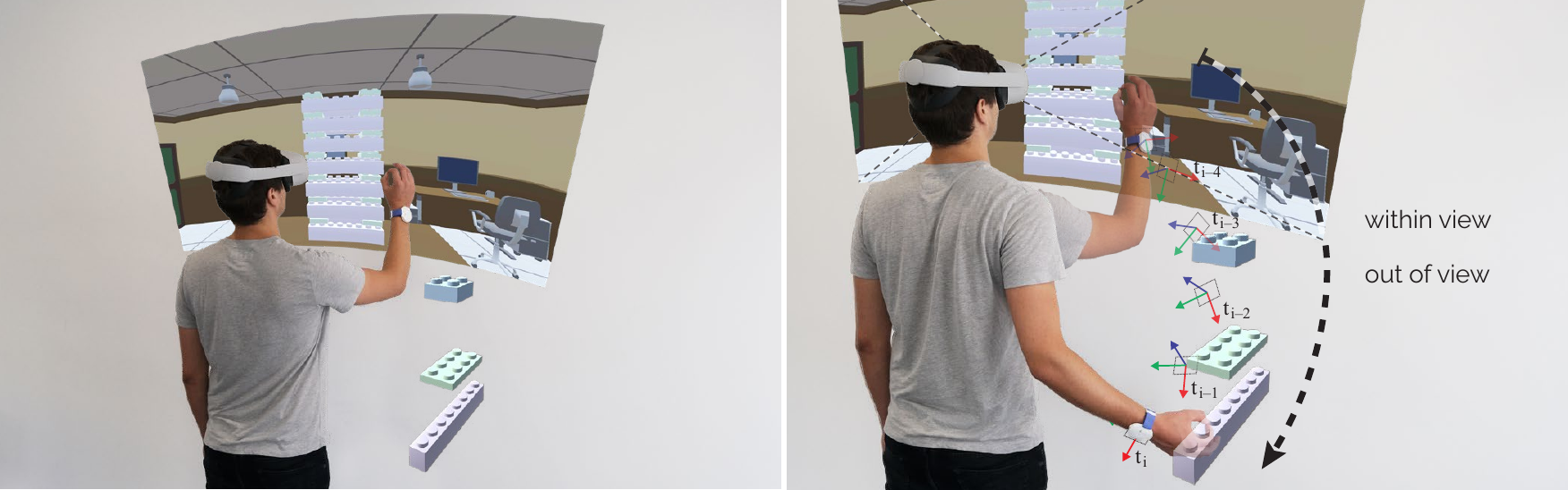}
    \caption{
\emph{\projname} is a wireless sensing method that complements existing virtual and augmented reality headsets to support hand tracking outside the field of view of the headset's cameras.
In this example app of our system, a participant is building a structure composed of blocks with different sizes.
When the user places an object onto the structure, his hand is tracked by the headset's cameras (\textit{left}).
As the user reaches for a block just to his right (\textit{right}), his hand leaves the field of view of the cameras.
Now, \projname\ continuously estimates the 6D position and orientation of the wrist using the 6-axis inertial measuring unit attached to the wrist during the time the hand stays outside the field of view.
    }
    \Description{The figure consists of two pictures of a user wearing an Oculus Quest 2 headset with HOOV attached to his right hand's wrist. In the left picture, the user interacts with building blocks in a tower structure within the field of view of the headset's cameras. In the right picture, the user grabs for another building block outside the cameras' field of view.}
    \label{fig:teaser}
\end{teaserfigure}

\maketitle

\section{Introduction}

Mixed Reality systems are increasingly designed to be standalone, both for Augmented Reality (AR) as well as Virtual Reality (VR) scenarios.
They typically integrate all necessary components for an immersive virtual experience in a user-worn headset.
These include a stereoscopic display, stereo sound, processors for rendering, as well as sensors such as inertial measurement units (IMUs) and cameras for inside-out head-pose tracking in world space.

Standalone systems are highly portable, allowing them to operate in mobile scenarios, but requiring them to make several trade-offs in their design.
Because their provided comfort strongly affects the user experience, size, form factor, and weight are the limiting factors in the implementation.
These considerations ultimately limit the available computing power, battery capacity, and display size, which confines the \emph{visible} field of view (FOV) for the wearer.
They also affect the number, type, and placement of integrated sensors, in particular the cameras that track the user's surroundings and, thus, the effective \emph{tracking} FOV.
In addition to scene perception, these embedded cameras increasingly deliver the signal for detecting user input through hand gestures and interaction with virtual and physical objects.
Therefore, the effective tracking FOV also determines the operational range for hand input on these devices.

In practice, today's headsets typically cover an operational range that slightly exceeds the user's visible FOV, thereby requiring all hand-object interaction to happen under \emph{visual control}.
While the operational range could be increased with additional headset cameras, which would incur additional needs for compute, power, and physical space inside the headset, certain areas around the body might still be outside the line of sight due to occlusion caused by clothing, hair, or other parts of the user's body.

As a result, today's systems implicitly require all interaction to occur in the visual FOV, which neglects most of the available space around the user for input.
We argue that this is a missed opportunity, as humans perceive more than 210\degree~\cite{visualfield}, allowing us to perform hand-object interactions even at the edge of our periphery.
Even without visual control, we routinely rely on proprioception to quickly place and retrieve physical objects in our vicinity~\cite{mine1997proprioception, yan2018eyesfree}---without requiring a turn of the head, which would slow down such motions.
Since such short-term use cases do not justify the cost of additional cameras, \emph{off-screen interaction} is not part of the interaction vocabulary on today's headsets.
And with mixed reality devices' ever-decreasing form factor, we do not expect future devices to substantially widen the tracking FOV.

In this paper, we introduce \emph{\projname}, a method to track interaction outside a headset's tracking field-of-view in immersive environments.
Using our novel data-driven inertial estimation pipeline, \projname\ complements current headsets by leveraging the continuous signals from a 6-axis IMU inside a wrist-worn band to estimate the user's current hand position outside the headset's tracking FOV.

\subsection{Outside field-of-view interaction}

\autoref{fig:teaser} shows an example application of our method.
Here, a user is building a structure composed of blocks with different sizes, placed by his side for easy access.
The user can pick up a building block by pinching it, dragging it to the tower, and dropping it at the desired location and with the hand-held orientation.
The building blocks by his side replenish, such that grabbing one immediately produces another one at the same location.

Having built up the tower so far, the user has internalized the location of building blocks by his side by now.
Naturally, as time progresses, he has to rely less and less on visual operation and, instead, simply reaches for one of the three locations to grab the corresponding piece.
This aptly supports his construction process as the complexity of the structure advances, allowing him to keep his visual attention fixated on the tower and the spot where he plans to place the next block, while reaching out to grab it.

The underlying implementation of this is powered by our method \projname.
When the user's hand leaves the headset's tracking FOV, \projname\ takes over the tracking and continuously provides the hand's current 3D position to the application.
For this, \projname\ processes the signals from the 3-axis accelerometer and 3-axis gyroscope sensor integrated inside the wrist-worn band as input.
From these, \projname\ estimates the wrist's current 3D orientation and feeds it into our novel temporal machine learning model, which estimates the current hand position.

\begin{figure}[t]
    \centering
    \includegraphics[width=\columnwidth]{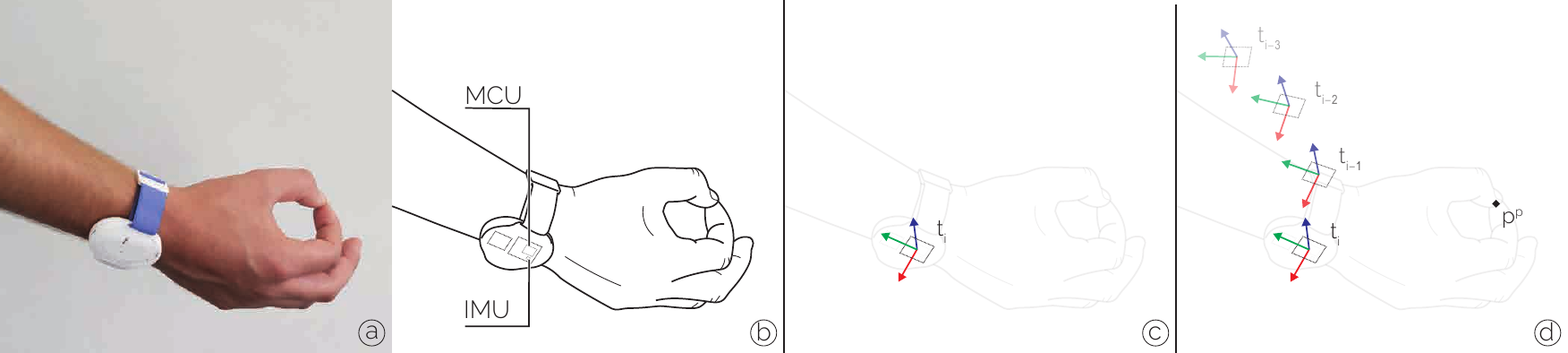}
    \caption{\projname's tracking pipeline to predict outside field-of-view hand positions.
    (a)~The wrist-worn band contains an embedded platform that houses (b)~a 3-axis accelerometer and a 3-axis gyroscope.
    (c)~From the signals of these inertial measurement units (IMU), \projname\ first estimates the wrist's 3D orientation and
    (d)~feeds the series of orientation estimates as input into our novel temporal machine learning model to predict the position of the hand.
    }
    \Description{The figure consists of four separate subfigures. Figure (a) shows a close-up picture of the user's wrist with the worn HOOV wristband. Figure (b) shows a line drawing of the same picture including the annotated IMU and CPU components affixed to the PCB contained within HOOV's case. Figure (c) illustrates the orientation of the wrist through a coordinate system anchored at the PCB board in the line drawing. Figure (d) shows the change in orientation of the wristband over time through a drawn sequence of coordinate systems.}
    \label{fig:method_preview}
\end{figure}

To register input commands from the user, \projname\ leverages the same IMU to detect hand gestures.
In \autoref{fig:teaser}, the user pinches to grab the intended building block, which \projname\ recognizes as a performed gesture in the signal stream.
\projname's pinch detection complements the interaction paradigm on today's VR platforms, in which pinching is commonly used either for direct or remote selection through a ray~\cite{POUPYREV199919, gazepinch}.

\subsubsection*{Performance Evaluation}

We conducted two studies to evaluate \projname's performance during outside field-of-view interaction in stationary scenarios.
Both studies consisted of three tasks each. 
In the first task, participants grabbed objects outside their FOV while facing forward, resembling our examples in \autoref{fig:teaser} but scaled up to 17~potential targets.
In the second task, participants placed virtual spheres into 1 of 17 discrete drop zones by their side.
In a final compound task, participants repeatedly switched between retrieving and placing objects outside their FOV, which constituted a compound and, thus, longer task.
In \studyone{}, participants' wrists were tracked by an 8-camera OptiTrack Prime\,13 system, permitting an analysis of the human ability to interact based on proprioception only, under nearly optimal tracking conditions.
Participants achieved a success rate of $\sim$91\% when grabbing and dropping objects off-screen.
In \studytwo{}, participants were operating exclusively using \projname's tracking when their hand left the headset's view.
They were successful in 86.18\%, 87.65\% and 86.18\% of cases for the grab, placement, and compound task respectively.

When considered as a target-agnostic 3D tracking system, \projname's simulated position estimates form a 3D trajectory that has a mean absolute error of 7.77\,cm from the OptiTrack's reference path in \studyone.
We observed that \projname's error was lowest within the first three seconds, covering the duration of most interactions outside the user's field of view.
In our simulations of lower tracking FOV in the headset (i.e., earlier moments at which the headset hands off tracking to \projname), we found that \projname's tracking error slightly increased, with a mean absolute error of 9.28\,cm in a simulated 120\degree\ tracking FOV and a 10.16\,cm error in a simulated 90\degree\ tracking FOV.
In \studytwo{}, the average error remained below 16\,cm during real-time interactive use over longer off-screen trajectories.
In its current implementation, \projname's inference pipeline runs in real-time on a desktop machine with an \textit{NVIDIA GeForce RTX 3090} GPU, requiring 4\,ms to predict a 3D hand pose, which amounts to a maximum update frequency of 250\,Hz.

In both studies, participants also completed a separate condition of the same tasks under visual control.
In this condition, the hand was only tracked by the headset, in our case an Oculus Quest~2, requiring participants to turn their head to keep the hand inside the headset's field of view.
While the average success rate rose to 95\% across all tasks in \studytwo{}, the average task completion time also increased, most notably by the compound task.
Due to the monotony and repetitiveness of the task, most participants pointed out the unfavorable need for turning their heads in this condition and the strain this puts on their neck over time, preferring interaction through \projname\ over interaction under visual control.

\subsubsection*{Applications}

\begin{figure*}
    \centering
    \includegraphics[width=\textwidth]{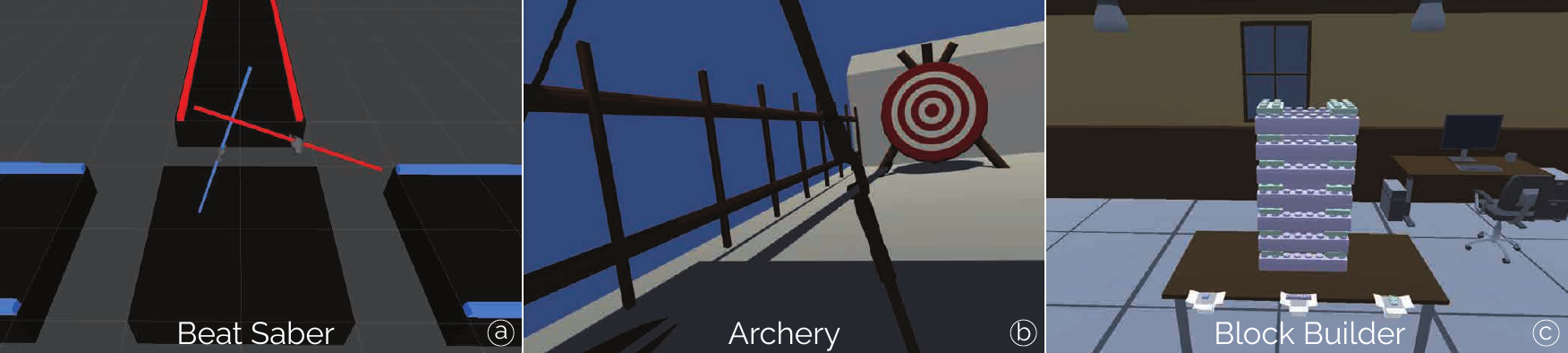}
    \vspace{-5mm}
    \caption{Our demo applications demonstrate the benefits of our sensing method for three different use cases: (a) a user plays an adapted version of Beat Saber with objects arriving from four different directions that require a wide tracking range to support the intended movements, (b) while shooting an arrow users pull the bowstring far into the back requiring hand tracking outside the headset cameras' FOV, and (c) a block builder app where users can reach for objects outside their FOV.}
    \Description{The figure shows three screenshots, one for each demo application. The left figure shows a screenshot of the beat saber app. Two double-bladed light sabers are held by a person standing on a black platform and waiting for virtual blocks to arrive from four orthogonal directions. The second picture (archery app) shows a tense bow with an arrow pointing towards an archery target. The third screenshot shows the environment of the block builder app. A set of building blocks are arranged to a tower on top of a brown table.}
    \vspace{-3mm}
    \label{fig:apps}
\end{figure*}

\projname\ enables several scenarios that are directly applicable and useful for immersive systems.
We show this at the example of three demo applications that benefit from the knowledge of the current hand position beyond the trackable FOV of the headset's cameras.
In the first application, users can interact with building blocks outside their FOV as illustrated in \autoref{fig:teaser} and \autoref{fig:apps}.
In the second scenario, \projname\ improves a \textit{Beat Saber} gaming experience by tracking quick arm movements that cross hardly tracked areas around the user's body.
Finally, \projname\ facilitates tracked arm movements of extended range as demonstrated in our third application, an archery game.

We conclude this paper with a discussion of our applications and the implications of our method for future immersive systems.

\subsection{Contributions}

In this paper, we contribute

\begin{itemize}[leftmargin=*]
    \item a tracking method that enables short-term interaction beyond the FOV of the cameras inside AR/VR headsets.
    Our method merely relies on the signals from a 6-axis IMU, integrated into our custom inertial sensing device placed on the user's wrist, which captures the information for outside-field-of-view 6D hand tracking and pinch detection and provides haptic feedback during interaction events.
    
    \item a novel temporal deep learning architecture that comprises a Transformer for the estimation of the current hand position from the IMU signals and the latest available output of the headset's visual hand tracking pipeline.
    
    \item a user study with 12 participants to investigate human proprioceptive abilities and \projname's tracking performance, where participants retrieved and placed spheres at 17 different target locations outside the visual and the camera's field of view.
    Using an 8-camera OptiTrack system, participants grabbed and placed the correct object $\sim$91\% of the time.
    In an offline simulation, they would have grabbed and placed the correct object in $\sim$85\% of all cases using \projname{}.
    Compared to a baseline with a commodity headset where participants grabbed and placed objects under visual control, their success rate increased to $\sim$94\% at the expense of 20\% slower task completion and repeated neck movement.
    
    \item a user study with 10 participants where participants completed the selection and placement tasks using \projname{} in real-time, achieving success rates of more than 86\%.
\end{itemize}

\section{Related Work}

\projname{} is related to body capture using worn sensors, inertial tracking, and interactions driven by spatial memory, proprioception, and kinesthesia. 

\subsection{Body capture using worn sensors}

Information about the user's body pose, especially the posture of the arm, enables VR and AR systems to correctly embody the user and to detect input through gestures.

In contrast to external body pose capture systems that include high-end commercial marker-based camera systems~\cite{optitrack}, depth and RGB~\cite{guler2018densepose, cao2017realtime, michel2017markerless} cameras as well as non-optical systems~\cite{zhao2018through, zhang2018wall++, northern}, body-worn systems enable tracking in mobile settings, and do not require the user to remain within a constraint area.
To receive a wider view of the user's body for tracking, cameras with wide angle-lenses were attached to various parts of the body including the wrist~\cite{10.1145/3379337.3415897}, the feet~\cite{bailly2012shoesense}, and the chest~\cite{hwang2020monoeye, jiang2017seeing} or alternatively integrated into controllers~\cite{ahuja2022controllerpose} or using a suspension on the headset~\cite{10.1145/3332165.3347889, 10.1145/2980179.2980235} further away from the user's body.
However, these solutions tend to suffer from occlusion and are often obtrusive to wear.
Alternative approaches directly estimate the full-body pose based solely on the available temporal motion information of the user's head and hands~\cite{jiang2022avatarposer, ahuja2021coolmoves, winkler2022questsim}.
Moreover, various specialized mobile hand-held or body-worn systems have been built for tracking that make use of sensing modalities such as magnetic~\cite{chen2016finexus}, mechanical~\cite{metamotion}, or acoustic~\cite{Jin2015,vlasic2007practical} sensing.

Prior research has also shown that body-attached IMUs can be used in a standalone system for tracking human pose.
Depending on the number of body locations that are instrumented, the problem varies in difficulty.
Sparse Inertial Poser~\cite{von2017sparse} estimates the 3D human pose based on 6 IMU sensors attached to the wrists, lower legs, the back and the head using a joint optimization framework incorporating anthropometric constraints.
Deep Inertial poser~\cite{DIP:SIGGRAPHAsia:2018} trains a recurrent neural network (RNN) for this task.
Transpose~\cite{yi2021transpose} and PIP~\cite{yi2022physical} further improve on the predicted output by incorporating joint-specific loss terms and physic-aware motion optimizers.

\subsection{IMU-based arm and wrist tracking}
Instead of tracking the full body, \citeauthor{li2020mobile} attached three IMU sensors to the hand, forearm and upperarm to track a user's arm pose through solving inverse kinematics~\cite{li2020mobile}.
Closely related to our work, \citeauthor{shen2016smartwatch} presented an online and an offline tracking method to estimate the posture of the arm based on a single wrist-worn 9-axis IMU integrated within a smartwatch~\cite{shen2016smartwatch}. 
The method finds an optimal path through a set of potential candidate arm postures that are retrieved from a dictionary based on the IMU-estimated wrist orientation at each time step.
The offline version uses Viterbi decoding to estimate an optimal tracking path, and achieves a medium tracking error of 9.2\,cm for the wrist.
However, since it occurs a complexity of $\mathcal{O}(N^{3}T)$, it is not suitable for real time-tracking.
A simpler online version that estimates the arm position from the candidate postures at a single step using a frequency-weighted average achieves a median tracking error of 13.3\,cm.
\citeauthor{liu2019poster}~\cite{liu2019poster} propose to improve the computation based on Hidden Markov Model state reorganization, achieving an accuracy of 12.94\,cm.
With an updated algorithm including a Particle Filter, a median error of 8.8\,cm was reported in a static setting~\cite{shen2018closing}.

Similarly, \citeauthor{wei2021real} propose an RNN architecture to track the arm posture from a 6-axis IMU assuming a fixed shoulder position~\cite{wei2021real}.
They report a median error of 15.4\,cm and a MAE of 16.4\,cm for the wrist in a leave-one-subject-out evaluation scheme.
LimbMotion uses an additional edge device for acoustic ranging to estimate the arm posture with a median wrist error of 8.9\,cm~\cite{zhou2019limbmotion}.

Compared to the previous approaches, our method does not aim to support stand-alone arm tracking based on a 6-axis IMU but to support a visual tracking system for short periods of time where the hand moves outside the cameras field of view.
This allows us to correct for drift, especially in the yaw direction which is more pronounced due to the missing magnetometer, as soon as the hand is visible by the headset's cameras.
Moreover, our method does not assume a fixed shoulder position.

\subsection{Spatial Memory, Proprioception, and Kinesthesia}
The human ability to operate without visual control is mainly supported by the following factors: spatial memory \cite{scarr2013spatialmemory}, proprioception \cite{cockburn2011airpointing}, and kinesthesia \cite{li2010access}. 
Spatial memory refers to the part of memory responsible for recording the position and spatial relations between objects \cite{gustafson2010imaginaryinterfaces, baddeley1992workingmemory}. 
Proprioception refers to the sense of position and orientation of one's body parts with respect to each other \cite{cockburn2011airpointing, sherrington1907proprioception}.
Kinesthesia, which is often used interchangeably with proprioception, refers to the perception of one's body movements and motions \cite{li2010access}. In prior research, there is substantial work focused on characterizing the aforementioned cognitive and perceptual abilities of people \cite{biggs2002haptic}. 
Gutwin et al. \cite{cockburn2017docnav} and Cockburn and McKenzie \cite{cockburn20022D3D}, for instance, studied people's capabilities of building spatial memory in 2D and 3D spaces. 
Hocherman \cite{hocherman1993proprioceptive}, Soechting and Flanders \cite{soechting1989sensorimotor}, and Medendrop et al. \cite{medendorp1999pointing} examined the extent to which people can rely on their proprioception to perform target selections. 
Andrade and Meudell \cite{andrade1993spatial} and Postma and De Haan \cite{postma1996memory} showed that people are capable of learning spatial locations without paying particular attention to them. 

Within the human-computer interaction community, it is generally acknowledged that spatial memory, proprioception, and kinesthesia can be exploited to support rich and efficient interactions \cite{scarr2013spatialmemory, mine1997proprioception}.
Li et al.'s Virtual Shelves technique \cite{li2009shelves} leveraged users' kinesthetic memory for triggering programmable shortcuts.
Using Virtual Shelves, users can select shortcuts by pointing a spatially-aware device at 28 different locations in front of themselves.
Yan et al. \cite{yan2018eyesfree} build on top of this work and experimentally studied eye-free target acquisition in VR including the additional aspect of user comfort. 
Cockburn et al. \cite{cockburn2011airpointing} presented a design space exploration for the interaction termed "air pointing". 
Gustafson et al.'s Imaginary Interfaces \cite{gustafson2010imaginaryinterfaces} sought to enable bi-manual empty-handed interactions without visual feedback by relying on users' short-term memory.
Imaginary phone \cite{gustafson2011imaginaryphone} demonstrated the potential of transferring users' spatial memory from a familiar physical device to operating an "imaginary" equivalence on their palm. 
Additional works explored proprioception-driven interactions for menu usage~\cite{uddin2016handmark}, allowing information access via a hip-attached touch device~\cite{dobbelstein2015belt}, supporting mobile phone access for visually impaired users~\cite{li2010access}, and enabling interaction with the back of a headset \cite{gugenheimer2016facetouch}.

Also closely related to \projname{}, many works support users in proprioceptively performing interactions through haptics. 
\citeauthor{kim2016assisting}~\cite{kim2016assisting} guided visually-impaired users in target selection on a large wall-mounted display with vibrotactile feedback. 
Barham et al.'s CAVIAR device~\cite{barham2012CAVIAR} similarly used vibrotactile actuators to guide users' hands with continuous stimuli. 
Vo and Brewster ~\cite{vo2015touchinginvisible} studied ultrasonic haptic feedback to support spatial localization.

HOOV ultimately aims to enable the aforementioned spatial memory- and proprioception-driven interactions by expanding the tracking space of current VR devices for hand-object interactions. 
We focus on addressing the technical challenge of performing the sort of eye-free interactions explored by Yan et al. \cite{yan2018eyesfree} when the available headset tracking field-of-view is limited. 
We further support users in performing eye-free interactions with haptic feedback via an actuator integrated into our wrist-worn device.

\section{\projname\ Method: Estimating pinch positions from observations of wrist orientations}
\begin{figure*}[t]
    \centering
    \includegraphics[width=\textwidth]{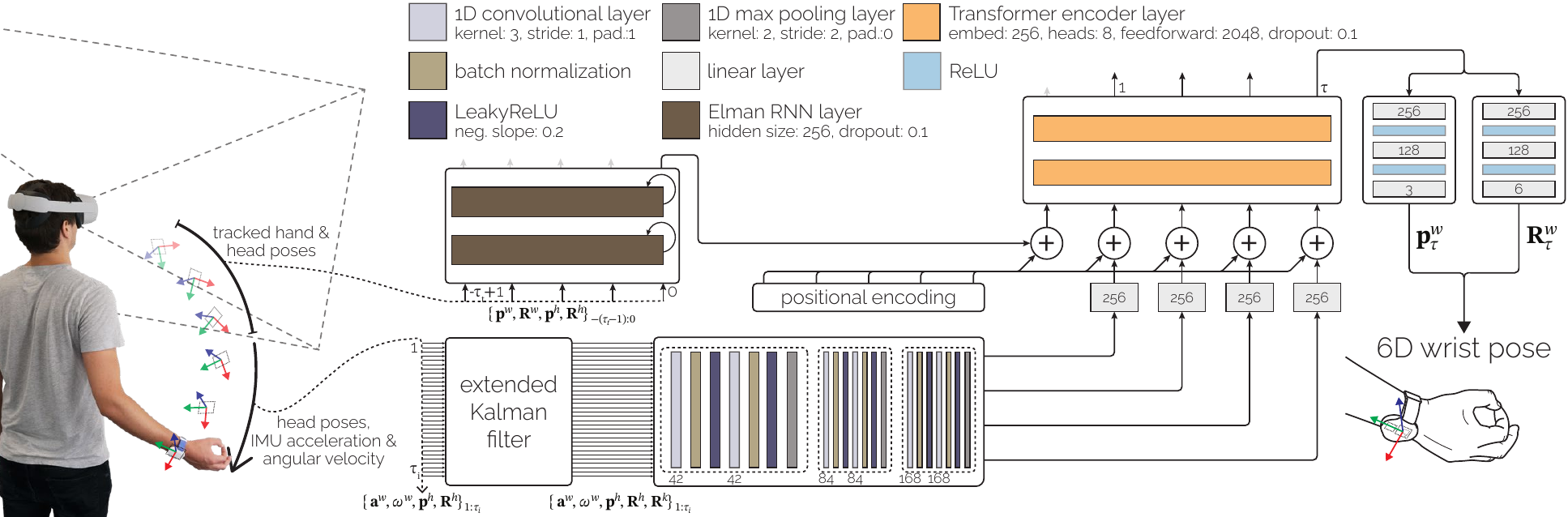}
    \caption{\projname's wrist pose estimation pipeline. A 2-layer RNN receives the last 5 available hand poses from the Oculus Quest 2 as input. The inertial motion data captured since the point where the hand left the FOV of the headset's cameras is downsampled by a factor of 8. The inertial data is then transformed by a linear embedding layer and appended to the output of the RNN. A Transformer processes the sequence before each sample is mapped to a position and an orientation.}
    \Description{The figure illustrates HOOV's wrist pose estimation pipeline. It depicts a user moving his hand outside the field of view of the headset's cameras. The last 5 available hand poses tracked by the Oculus Quest 2, when the hand was inside the field of view of the cameras, are fed to a 2-layer Elman RNN with a hidden layer size of 256. The last output becomes the first input to a 2-layer Transformer encoder with an embedding size of 256, a feedforward layer size of 2048, and 8 attention heads. The additional inputs to the Transformer are the output of a convolutional neural network module that downsamples the tracked head poses of the Oculus Quest 2 and the captured inertial data at the wrist when the hand is outside the field of view by a factor of 8. The inertial data is augmented by the wrist orientation estimated through an extended Kalman filter. The Transformer module computes an output vector for each downsampled inertial feature vector. Based on this, two sequences of linear layers predict an orientation and position estimate for the wrist pose respectively.}
    \label{fig:pipeline}
\end{figure*}

We now introduce our method that enables the tracking of the wrist position, $\mathbf{p}^{w} \in \mathbb{R}^{3}$, and orientation, $\mathbf{R}^{w} \in SO(3)$, based on the captured acceleration, $\mathbf{a}^{w} \in \mathbb{R}^{3}$, and angular velocity signals, $\mathbf{\omega}^{w} \in \mathbb{R}^{3}$, of a 6-axis IMU placed at the user's wrist.
Besides the wrist-worn band's motion signals, \projname\ also receives as input the headset's current position, $\mathbf{p}^{h} \in \mathbb{R}^{3}$, and orientation, $\mathbf{R}^{h} \in SO(3)$, as well as the sequence of $\tau_{t}$ last available head and hand poses $\{\mathbf{p}^{w}, \mathbf{R}^{w}\, \mathbf{p}^{h}, \mathbf{R}^{h}\}_{-(\tau_{t}-1):0}$ before the hand has left the tracking FOV of the headset.
We describe this mapping $f$ by
\begin{equation*}
    \{\mathbf{p}^{w}, \mathbf{R}^{w}\}_{1:\tau} = f(\{\mathbf{a}^{w},\mathbf{\omega}^{w},\mathbf{p}^{h}, \mathbf{R}^{h}\}_{1:\tau_{i}}, \{\mathbf{p}^{w}, \mathbf{R}^{w}, \mathbf{p}^{h}, \mathbf{R}^{h}\}_{-(\tau_{t}-1):0}),
\end{equation*}

where $\tau$ matches the number of considered time steps predicted since the hand has left the tracking FOV, and $\tau_{i}$ is the number of inputs sampled in the same period.

This is a challenging problem since we only receive the noisy observations of the relative motions of the wrist whose position lies in a 5-DoF space given a static shoulder position (3 DoF through shoulder joint and 2 DoF through lower arm)~\cite{shen2018closing} and a 7-DoF space given a constant neck position (additional 2 DoF from clavicle) as input.
However, the set of possible wrist poses is severely constrained given the knowledge of the forearm's orientation and by incorporating the anatomical constraints of the individual human joints~\cite{shen2016smartwatch, shen2018closing}.
While orientation estimation from a 6-axis IMU suffers from yaw drift due to the lack of a magnetometer, we demonstrate that a learning-based method can correct for this drift for short-term out-of-view interactions in static settings.
Our method does this by estimating the most likely arm trajectory from an implicitly acquired approximated distribution of previously seen trajectory observations.

In addition to tracking the 3D positions of the wrist, \projname\ detects input commands from the wearer.
We continuously process the signals of the accelerometer for characteristic patterns to detect gestures, such as pinching or fist clenching.
Using the build-in haptic actuator, our prototype is capable of rendering haptic feedback to the user in response via a knocking sensation.

\subsection{Inertial 6D hand pose estimation}

\autoref{fig:pipeline} shows an overview of our inertial 6D hand pose estimation pipeline.

\subsubsection*{Input and output representation}
The input consists of the accelerometer and gyroscope signals from the IMU within the wrist-worn band as well as the current head position and orientation estimated by the inside-out tracking of the VR headset. 
The spacing between the samples of tracked head positions is matched to the sampling interval of the IMU through cubic interpolation.
We further estimate an initial orientation of the wrist by directly applying an extended Kalman Filter~\cite{ekf} to the gyroscope and accelerometer signals~\cite{hartikainen2011optimal, sabatini2011kalman}.
We convert all orientations to their 6D representation to ensure continuity~\cite{zhou2019continuity}.
For each time step, we concatenate the acceleration, angular velocity, head position and orientation, and the output of the extended Kalman Filter $\mathbf{R}^{k}$. 
This input sequence, $\mathbf{X} \in \mathbb{R}^{\tau_{i} \times 21}$, starts from the moment when the hand leaves the trackable FOV of the headset.
We obtain another input sequence $\mathbf{S} \in \mathbb{R}^{\tau_{t} \times 18}$ of the last $\tau_{t}$ head and hand poses that were tracked by the headset before the hand left the tracking FOV.

Based on this input, our estimator directly predicts wrist position $\mathbf{p}^{w} \in \mathbb{R}^{3}$ and rotation $\mathbf{R}^{w} \in \mathbb{R}^{6}$ within the world.
Using the orientation of the wrist, we produce a more refined estimate of the pinch position, $\mathbf{p}^{p} \in \mathbb{R}^{3}$, by applying an offset of 15\,cm to the wrist position,
\begin{equation*}
    \mathbf{p}^{p} = \Tilde{\mathbf{R}}^{w}(0, -0.15, 0)^{T} + \mathbf{p}^{w},
\end{equation*}
where $\Tilde{\mathbf{R}}^{w}$ is the rotation matrix corresponding to $\mathbf{R}^{w}$ and the y-axis of the right-handed local coordinate system centered at the wrist points along the forearm to the shoulder with the z-axis pointing downwards through the palm.
We took this option because we expected position estimates to be better than representations that encode the rotations for each joint along the kinematic tree from the head to the wrist.
Since the interaction happens outside the user's visible FOV, visualization artifacts due to varying bone lengths are of limited concern.

\subsubsection*{Network architecture}

\projname\ relies on a neural network to approximate the mapping $f$.
The architecture of the network consists of a downsampling module that receives the input sequence $\mathbf{X}$ containing the inertial motion information as input and reduces the sequence along the temporal axis by a factor of 8.
The module consists of three blocks that each reduce the signal's temporal resolution by a factor of 2 and consist of two convolutional layers with a kernel of size 3 followed by a max pooling layer.
The samples of the remaining features are converted to linear embeddings that are fed to a Transformer consisting of two encoder layers.
At the start of the sequence, we add an initial token that embeds the information from the headset-tracked hand and head poses $\mathbf{S}$, extracted through a 2-layer Elman RNN.
We apply a sinusoidal positional encoding to the input of the Transformer~\cite{vaswani2017attention}, and avoid that future samples are attended to for the estimation of any output samples by applying a corresponding mask to the sequence.

For each output sample of the Transformer corresponding to the down-sampled features from the inertial input, we predict the position and orientation of the wrist using a series of fully-connected layers.
This design supports sequences of variable length.

\subsubsection*{Loss function}

Our loss function, 
\begin{align*}
   \mathcal{L} = \mathcal{L}_p + \mathcal{L}_R = \sum_{t=1}^{\tau} (|\hat{\mathbf{p}}^{w}_{t}-\mathbf{p}^{w}_{t}| + |\hat{\mathbf{R}}^{w}_{t}-\mathbf{R}^{w}_{t}|),
\end{align*}
consists of two terms, where $\mathcal{L}_p$ penalizes the positional and $\mathcal{L}_R$ the rotational offset using the L1 loss function, and $\mathbf{p}^{w}_{t}$ and $\mathbf{R}^{w}_{t}$ are the ground-truth and $\hat{\mathbf{p}}^{w}_{t}$ and $\hat{\mathbf{R}}^{w}_{t}$ the predicted position and orientation of the wrist at time $t$ respectively.

\subsection{Pinch detection}
\label{sec:pinchdetect}
To detect pinch events, we threshold the running rate-of-change score $c$, which accumulates the absolute change in the acceleration signals $\mathbf{a}^{w}$ captured by the IMU at the wrist across time~\cite{meier2021tapid, streli2022taptype},

\begin{equation*}
c_{t} = \tfrac{1}{D}c_{t-1} + \lvert\lVert\mathbf{a}^{w}_{t}\rVert_{2}-\lVert\mathbf{a}_{t-1}^{w}\rVert_{2}\rvert.
\label{eq:RCS}
\end{equation*}

Pinch events cause sudden changes in ${a}^{w}_{t}$ that lead to a strong increase $c_{t}$, and thus, can be detected through a threshold.
The exponential reduction factor $D$ attenuates past accumulations.

\section{Implementation}

\projname\ is implemented to run as a real-time interactive tracking system on an 8-core
Intel Core i7-9700K CPU at 3.60\,GHz with an NVIDIA GeForce RTX 3090 GPU.
\projname\ consists of three components: 1) a wrist-worn sensing platform, 2) a virtual reality interface, and 3) a central control and sensor fusion unit that handles the communication to the wristband hardware and the virtual reality headset as well as the estimation of the hand pose outside the FOV.

\subsection{Hardware}

\begin{figure}[t]
    \centering
    \includegraphics[width=\columnwidth]{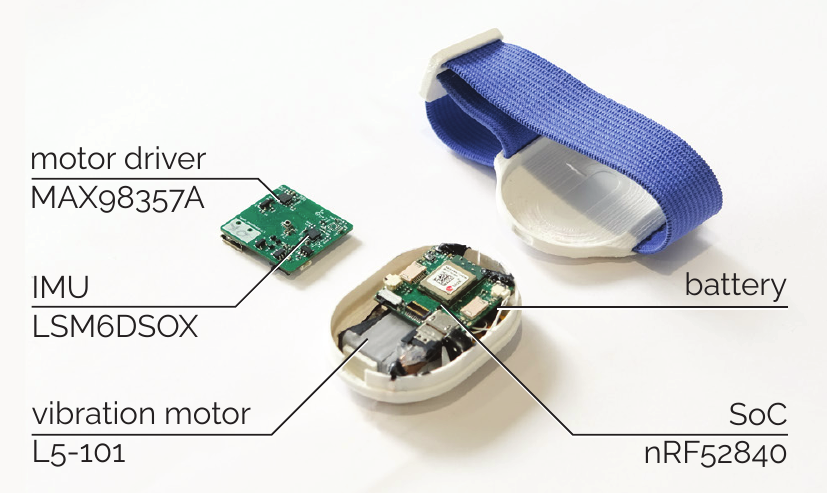}
    \caption{\projname's wristband integrates an electronics platform with a 6-axis IMU and a System-on-a-Chip within a 3D-printed case. The case also contains a battery and a vibration motor to provide haptic feedback to the user.}
    \Description{The figure shows a picture of the HOOV wristband with the white 3D-printed case opened. The case is attached to a flexible blue strap that the user pulls over their wrist. The case contains the electronics platform. The components of the electronics platform are labeled on the figure and include the motor driver (MAX98357A), the IMU (LSM6DSOX), the vibration motor (L5-101), the SoC (NRF52840), and the battery.}
    \label{fig:hardware}
\end{figure}

We custom-designed an embedded platform for \projname\ for sensing and actuation during interaction in mid-air.
As shown in \autoref{fig:hardware}, our electronics platform centers around a System-on-a-Chip
(NRF52840, Nordic Semiconductors) that samples a 6-axis IMU (LSM6DSOX, STMircroelectronics) at 427\,Hz. 
Our prototype streams the inertial data to a PC, either wirelessly over BLE or through a wired serial interface.
The prototype integrates a separate board that embeds an audio amplifier (MAX98357A, Maxim Integrated) to drive a Lofelt~L5 actuator to produce haptic feedback. 
All components of our prototype, including a battery, are housed in a 3D-printed case, which attaches to the wrist using a strap.

\subsection{Virtual Reality}

We implemented our VR application in Unity~2021 for the Oculus Quest~2 VR headset. 
Our application communicates with our outside-field-of-view tracking pipeline via a web socket. 
As tracking source, our VR environment receives the hand pose computed by the headset, sampled at $\sim70$\,Hz when the user's hand is tracked, and the estimates produced by \projname\ otherwise.
Input events are triggered through \projname's pinch detection algorithm in both states.

\subsection{Sensor fusion and outside FOV tracking}

The main control unit of our implementation is a state machine that moves between states depending on the current tracking status of the VR headset.
During the outside FOV state, the headset forwards the received head poses and the inertial input data together with the last 5 available hand and head poses from the VR headset to the deep learning-based hand pose estimation pipeline.

We implemented the main control unit in Python~3.8.
The program runs across multiple cores to handle the communication with the sensing hardware and the processing of the corresponding signals, and to set the current hand pose for the VR interface.

Upon detecting a pinch gesture using an exponential reduction factor $D$ of 1.07 and a threshold of 4.9\,m/s$^{2}$, our system triggers an event inside the VR environment and sends a command to the motor driver to activate the haptic feedback once.

\projname's network is implemented in \textit{PyTorch} and has 4,408,199 trainable parameters.
We use the Kaiming initialization~\cite{he2015delving}, the Adam optimizer~\cite{kingma2014adam} with a learning rate of $10^{-4}$, and a batch size of 16 where we randomly group sequences of equal length.

\section{\studyone{}: Evaluation where participants interacted based on \optitrack\ tracking}

To evaluate \projname's potential, we conducted a user study in which participants placed and retrieved objects outside their FOV.
\studyone{} served two purposes: (1) It allowed us to quantify \projname's accuracy of tracking wrist positions outside the headset's tracking FOV. (2) We could evaluate how quickly and accurately participants selected and placed objects outside their field of view compared to performing the same task under visual control by turning their heads.
This quantified the effect of operating under proprioceptive control.
To evaluate the upper bound of our approach, we conducted this first evaluation while participants interacted based on OptiTrack tracking to exclude the impact of tracking errors on participants' behavior.
(See Section~\ref{sec:online_evaluation} for our real-time evaluation of \hoov\ for the same tasks (\studytwo{}).)

\begin{figure*}[t]
    \centering
    \includegraphics[width=\textwidth]{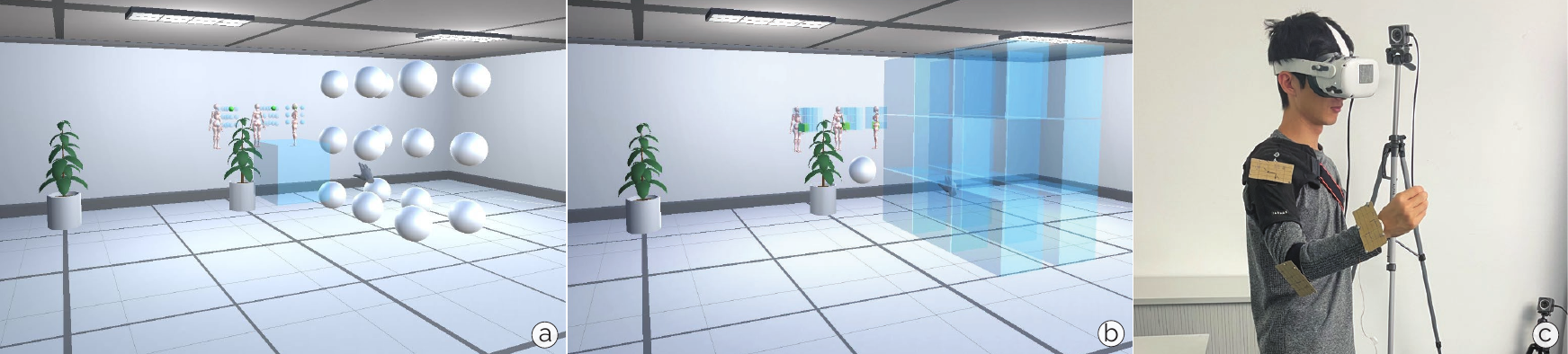}
    \caption{Study environment. Participants stood next to the set of spheres, focusing their gaze towards the wall onto a highlighted position, such that the spheres were to their right outside the headset's field of view.
    (a)~In the first task \grab{}, participants grabbed one of the 17 spheres outside their FOV and placed it in front of them into the blue cube.
    (b)~In the second task \drop{}, participants grabbed the sphere in front and dropped it into one of the 17 drop zones outside their FOV.
    In the third task \compound, participants dropped a sphere into a drop zone outside their FOV (b), and then retrieved another sphere outside their FOV (a).
    (c)~Participants wore a Quest~2 headset and our apparatus tracked their shoulders, elbows, and wrists using rigid-body markers and a high-resolution OptiTrack system.}
    \Description{The figure consists of three subfigures. The left figure (a) shows a screenshot of the virtual study environment for the grab condition. There are 17 spheres next to right side of the participant arranged in a grid of 3 rows, 3 columns, and 2 layers in the lateral direction. There is a blue rectangular drop zone in front of the participant as well as a down-sized illustration of the task environment, which highlights the sphere outside their FOV to pick next. Figure (b) shows a screenshot of the virtual study environment for the drop condition. There are 17 drop zones next to the right side of the participant. In the front is the sphere that the participant should place in the target drop zone highlighted in the down-sized illustration of the task environment. Figure (c) shows a participant wearing the Oculus Quest 2 headset, the HOOV wristband, and the rigid bodies made out of cardboard attached to the participant's shoulder and elbow as well as the HOOV wristband.}
    \label{fig:task_env}
\end{figure*}

\subsection{Study design}
\label{sec:studydesign}
\subsubsection{Apparatus}
Our experiment apparatus consisted of an Oculus Quest~2 VR headset, a \projname\ wristband as described above, a computer for logging, and an optical motion capture system.

\paragraph{VR headset}
Throughout the study, participants wore a Quest~2 VR headset with integrated visual hand tracking.
For the study setup, we implemented a VR environment featuring objects, placement zones, and task protocols in Unity~2021.
Participants saw a visualization of their right hand when it was within their FOV.

\paragraph{\projname's IMU streams and haptic feedback}
Participants wore a \projname\ wristband on their right arm, which continuously streamed the data from the IMU to a PC for processing.
From the continuous stream of accelerations, our apparatus detected pinch events as described in Section~\ref{sec:pinchdetect}.
To eliminate the impact of wireless connectivity onto participants' performance, we connected the wristband through thin and flexible magnet wires to the PC for power and reliable serial communication in this study.

\paragraph{Motion capture for 3D wrist and head trajectories}
To compare the accuracy of \projname's estimated wrist position to a ground-truth baseline, we tracked rigid-body markers attached to the participant's headset, shoulder, elbow, and wristband using an 8-camera OptiTrack system with sub-mm accuracy to obtain positions and rotations at a sampling frequency of 240\,Hz.
The rigid bodies are made out of cardboard that we attached to worn elbow and shoulder pads, and the HOOV wristband (\autoref{fig:task_env}).
To track the headset, we directly attached four 9-mm tracking markers to the device.

An experimenter ensured a consistent placement of the rigid bodies across participants ahead of the experiment and verified throughout that the markers did not slip.
We calibrated the OptiTrack to the coordinate system of the VR headset through two pre-defined calibration points in the physical space that we marked with a controller in the virtual environment.

\subsubsection{Participants}

We recruited 12 participants from our local university (3 female, 9 male, ages 22--35, mean=25.7 years). 
Using the OptiTrack, we measured the distance between the worn headset and the floor as well as the distance between the neck and the wrist while participants stood in a T-pose.
The average headset distance from the ground was 173\,cm (SD=12\,cm, min=149\,cm, max=188\,cm), and arm lengths ranged between 62\,cm and 80\,cm (mean=72\,cm, SD=6\,cm). 
Participants self-reported their prior experience with VR technology on a 5-point Likert scale (from 1--never to 5--more than 20 times).
Participants' ratings ranged between 1 and 5 and their median prior experience was 3.
Each participant received a small gratuity for their time.

\subsubsection{Task}
The study consisted of three grab-and-place tasks.
Throughout the experiment, participants stood at a fixed point that was highlighted on the floor using tape, such that the experimenter could verify their position.

In the first task \grab\ (\autoref{fig:task_env}), participants stood next to a set of 17 spherical target objects outside their FOV in the virtual environment, grabbed an intended sphere, and placed it into the blue dropzone in front of them.
Participants saw a down-sized illustration of the task environment in front of them, which highlighted the sphere to pick next.
Participants grabbed a sphere using a pinch gesture, selecting the sphere that was closest to their right hand when in range.
Participants then moved it towards the dropzone and released it using a second pinch.

In the second task \drop, participants picked up a spherical target in front of them, and placed it within one of 17 drop zones to their right outside their FOV.
Again, a set of three matching illustrations highlighted the task's target drop zone next to them.
Participants first grabbed the sphere using a pinch gesture, before moving it to the target drop zone.
After pinching again, the apparatus placed the sphere in the drop zone that was closest to the hand's position at the time of the second pinch.

The third task \compound\ combined Tasks~2 and~1.
First, participants grabbed the sphere within their field of view in front of them and dropped it into one of the 17 drop zones outside their field of view as instructed~(\autoref{fig:task_env}b).
Then, they were instructed to grab one of the 17 spheres from outside their field of view and release it within the drop zone in front of them~(\autoref{fig:task_env}a).

As shown in \autoref{fig:task_env}, the spheres and drop zones were arranged in a grid of 3~rows, 3~columns, and 2~layers in the lateral direction of the participant.
We excluded the closer center location behind the participant, because reaching it would require contorting one's arm in an uncomfortable pose without turning one's upper body.
All other targets outside the FOV were convenient to reach.

The spacing of the targets and drop zones remained static throughout the study, and was adjusted at the beginning of the experiment to each participant's eye level and arm length to ensure all targets could be reached.
The spacing in the lateral direction between the two layers was equal to half the length of the participant's arm.
The spacing between objects and drop zones in the sagittal plane was equal to half the distance between the neck and the wrist of the participant when performing a T-pose, corresponding to an angular spacing of around 30\degree\ between the drop zones and targets further away from the participant.
The target and drop zone grid was centered at the height of the participant's shoulder, and placed at a distance so that the participant's hand would reach all elements of the second layer when elongated.

\subsubsection{Conditions}
Participants performed the three tasks in two conditions.

In the \oculus\ condition, the participant's hand was solely tracked by the Quest~2 headset.
Thus, participants needed to turn their heads during the task and follow their hands to ensure that the headset tracked the hand's position when performing the task.

In the \optitrack\ condition, participants were instructed not to turn their head and to look straight forward.
An experimenter ensured this throughout by monitoring the head-mounted display through a secondary screen.
During this condition, the wrist position and orientation was provided to the VR environment by the OptiTrack motion capture system.

In both conditions, the apparatus detected pinch gestures for grabbing and releasing the spheres using the IMU inside the \projname\ wristband as in Section~\ref{sec:pinchdetect}.
Participants received haptic feedback for detected pinch events and were notified of erroneous selections and placements with audio feedback.

\subsubsection{Procedure}

The study started with a brief introduction of the tasks.
The experimenter then noted down the participant's age and gender.
Participants performed a T-pose to calibrate the system to their body size.
The study started with a training session of all conditions and tasks, followed by the experiment.

During training, participants performed 17 trials, one for each target location, for each task.
When using the \optitrack\ condition, participants received visual guidance for the \drop\ and \grab\ task through the down-sized visualization to allow them to refine their internal model of their hand position in 3-space.
The visualization highlighted the corresponding sphere or drop zone as soon as the participant's hand came into contact with it.
This online highlighting was exclusive to the training session and was not available during the actual experiment.

After training, participants performed all tasks in each of the conditions without guidance.
We counterbalanced the order of the task and conditions across participants.
Participants performed 34~trials for \drop\ and \grab, and 68~trials for the \compound\ task, where a single trial consisted of placing or retrieving a single target object.
For \drop\ and \grab, each drop zone and sphere was used as target twice with the order across targets randomized. The same drop zone or target sphere was not used in direct succession.
For the \compound\ task, we fixed the order of locations between directly successive \drop\ and \grab\ subtasks but randomized the order across combinations.
We ensured that each valid position was used twice as drop zone and target object location.
In total, participants completed 2 conditions $\times$ (34~\drop\ trials + 34~\grab\ + 68~\compound\ trials) = 272~trials in under 45 minutes.
Participants were instructed to complete the tasks as fast as possible while avoiding mistakes.

\subsubsection{Measures}
As dependent variables, our apparatus measured the time it took participants to complete each trial and the percentage of successfully completed trials.
A trial was successful when the participant placed the correct sphere in the correct drop zone.

\subsection{Comparing \projname\ with high-accuracy tracking from a motion capture system}
\label{sec:comparehoovopti}
Our study apparatus logged all signals and, thus, recorded a labeled dataset of IMU signals and ground-truth wrist and head positions from the motion capture system.

We used the recordings to train and test \projname's estimation network for an offline evaluation.
This allowed us to not only quantify \projname's performance given the specific configuration of the headset, but we could also simulate a variety of FOV configurations and, thus, moments at which \projname\ took over tracking from the headset due to our knowledge of the hand positions relative to the head pose at every time step.
The \optitrack\ condition thereby acts as the upper bound for participants' overall performance given the tracking system's high accuracy.
Using the OptiTrack data, we could evaluate participants' ability to interact with objects outside their FOV under near-perfect tracking conditions.
This enables us to decompose \projname{}'s error into contributions from participants' inherent limitation to perform proprioceptive interactions and limitations associated with our method itself.

\paragraph{Training and evaluation}
We evaluated our network in a 6-fold cross-validation where each fold used the data of two randomly selected participants for testing and the other 10~participants for training.
We further augmented our training data with the recordings from three additional participants that we acquired while piloting our study setup.
For training, we use all data from the \optitrack\ condition, including the trajectories recorded during participants' initial training sessions.
To further augment our dataset, we trained our network on sequences that simulate a horizontal FOV between 40\degree\ and 120\degree\ in steps of 5\degree.
We also generated training input based on the logged tracking information from the Quest~2.

We trained our network until convergence on validation data from a participant that was extracted and excluded from each training dataset.
We trained for at least 250,0000~iterations on an NVIDIA GeForce RTX 3090, which takes approximately 12~hours.
For the test set, we only use the trials of the \optitrack\ condition from the actual experiment.

We simulated and evaluated three headset tracking FOVs, including 90\degree\ (i.e., the visible FOV of the Quest~2), 120\degree\ (i.e., the approximate FOV of human binocular vision), and the actual tracking FOV of the Quest~2.
The latter exceeds 120\degree\ as we empirically determined through the headset's reported tracking state.

We evaluated \projname{}'s output in terms of mean absolute distance error (MAE) and mean angle of the difference rotation (MAD)~\cite{huynh2009metrics}, $\theta=2\arccos{\left(\abs{\hat{\mathbf{q}}^{w}\cdot\mathbf{q}^{w}}\right)}$, to the measurements reported by the motion capture system, averaged over the whole out-of-field-of-view tracking path.
Here, $\hat{\mathbf{q}}^{w}$ and $\mathbf{q}^{w}$ are the unit quaternions corresponding to the ground-truth orientation $\hat{\mathbf{R}}^{w}$ and predicted orientation $\mathbf{R}^{w}$ respectively, and $\cdot$ denotes the inner (or dot) product of vectors.
We also compare \projname{}'s success rates for the simulated FOV ranges to \oculus\ and \optitrack.

\subsection{Results}

\begin{figure}
    \centering
    \includegraphics[width=\columnwidth]{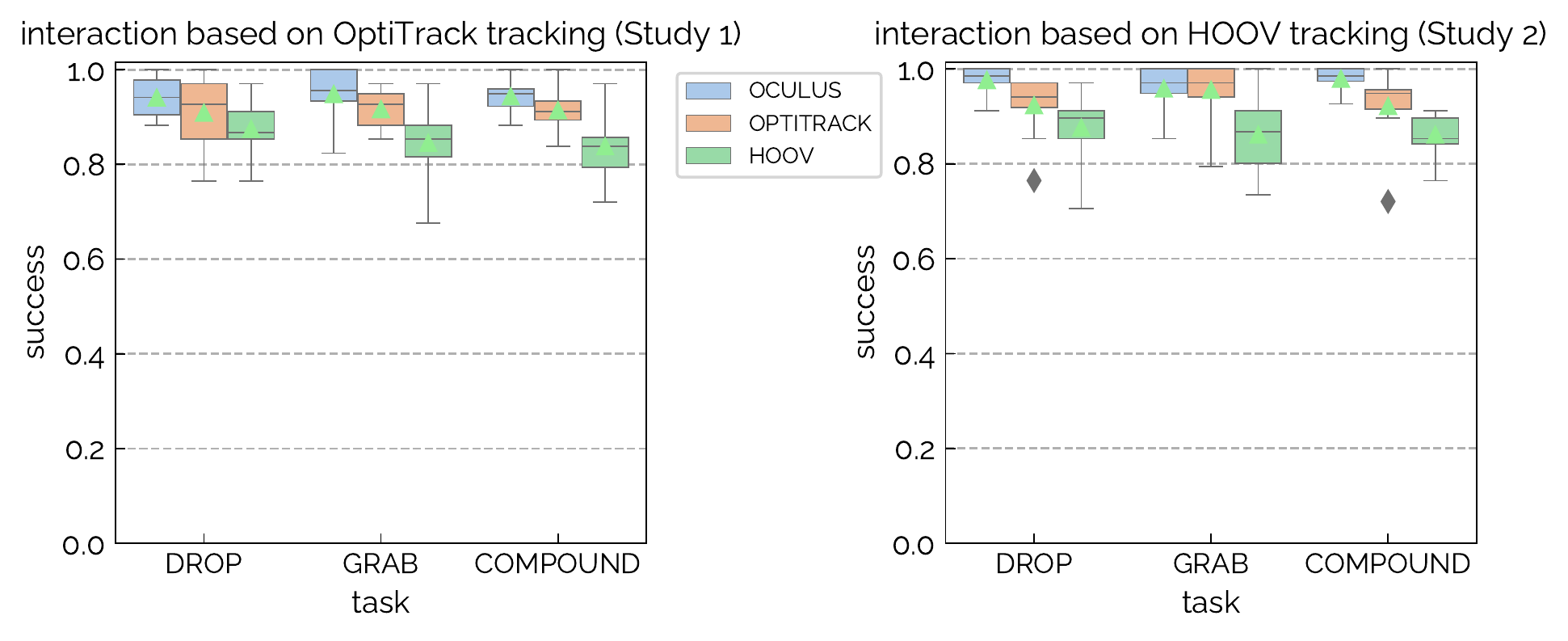}
    \vspace{-8mm}
    \caption{Mean success rates during the tasks for \studyone{} and \studytwo{}.
    (Left)~In \studyone{}, participants were most accurate when bringing targets into view (\oculus\ condition), with minor error rates when interacting outside their field of view (\optitrack\ condition) due to proprioceptive capabilities. With \hoov, participants reached 93\% to 96\% of this accuracy.
    (Right)~Error rates remained comparable during our online evaluation in \studytwo{} where \hoov\ was predicting input locations in real-time.}
    \Description{The figure consists of two box plots that illustrate the success rate for both Study 1 and Study 2 as described in Section 5.3.1 and Section 6.2.1.}
    \label{fig:boxplot_error}
\end{figure}

\begin{figure}
    \centering
    \includegraphics[width=\columnwidth]{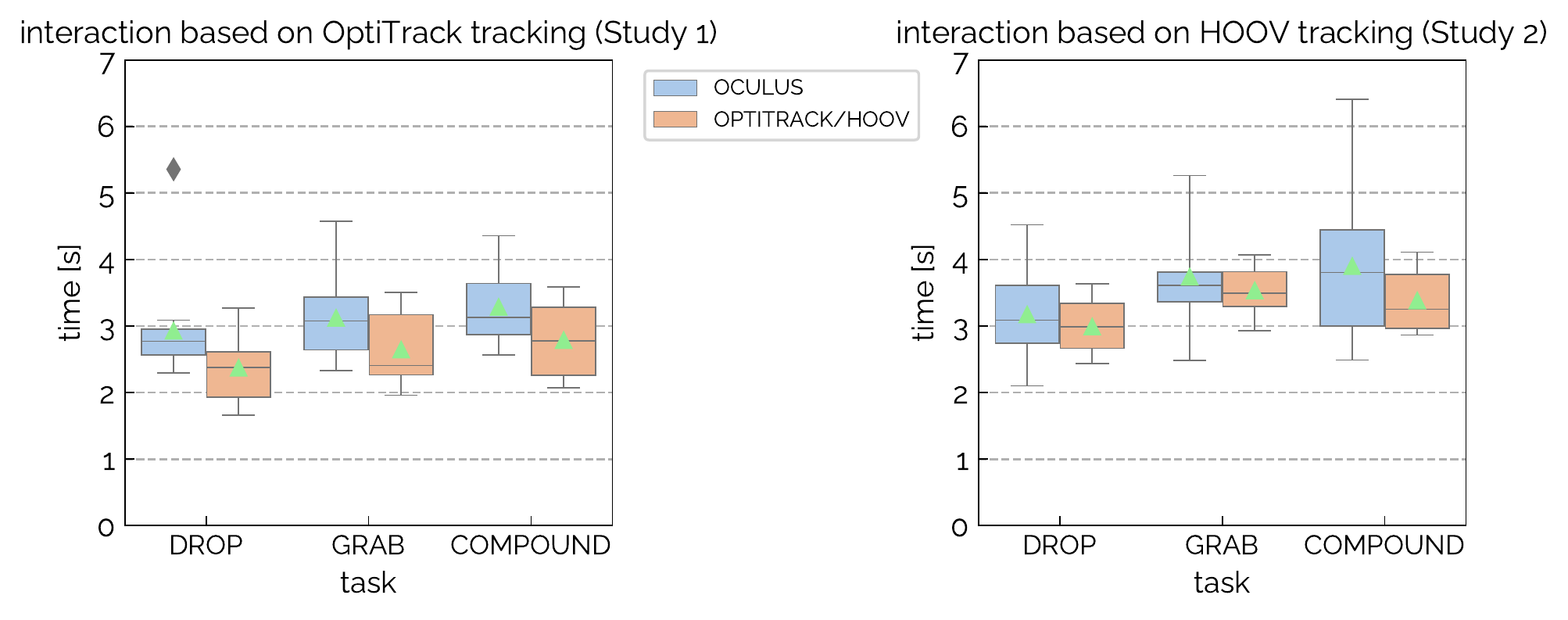}
    \vspace{-8mm}
    \caption{Mean trial duration of task completion for \studyone{} and \studytwo{}.
    (Left)~On average, mean completion time was longer in all three conditions when operating under visual control in \studyone{}.
    (Right)~In \studytwo{} with a fresh set of participants who were inexperienced VR users, mean completion time in the \oculus{} condition was again longer, most pronounced in the \compound\ task.}
    \Description{This figure consists of two box plots that illustrate the mean trial duration for both Study 1 and Study 2 as described in Section 5.3.2 and Section 6.2.2.}
    \label{fig:boxplot_time}
\end{figure}

\begin{table*}[t]
    \centering
    \caption{The table shows the tracking error of \projname\ with headsets of three different horizontal FOV ranges compared to a high-end motion capture system. The values are averaged across all participants. (±$\sigma$) is the standard deviation across participants. \textit{mean pos.} is the mean absolute error (MAE), \textit{median pos.} the median error and \textit{std. pos.} is the standard deviation within participants in terms of position in cm. \textit{mean rot.} is the mean angle of the difference rotation (MAD) and \textit{median rot.} the median angle of the difference rotation in \degree. \textit{std. rot.} is the standard deviation of the angle of the difference rotation within participants. \textit{mean DROP}, \textit{mean GRAB}, and \textit{mean COMP.} are the mean distances between the actual and \projname's estimate hand position at out-of-field-of-view release and grab events for the \drop, \grab, and \compound\ task, respectively.}
    \vspace{-2mm}
    \resizebox{\textwidth}{!}{
    \begin{tabular}{rccccccccc} \toprule
    \multicolumn{1}{c}{\textbf{FOV}} & \multicolumn{1}{l}{\textbf{mean pos.}} & \multicolumn{1}{l}{\textbf{median pos.}} & \multicolumn{1}{l}{\textbf{std. pos.}} & \multicolumn{1}{l}{\textbf{mean DROP}} & \multicolumn{1}{l}{\textbf{mean GRAB}} & \multicolumn{1}{l}{\textbf{mean COMP.}} & \multicolumn{1}{l}{\textbf{mean rot.}} & \multicolumn{1}{l}{\textbf{median rot.}} & \multicolumn{1}{l}{\textbf{std. rot.}} \\ \hline
    \textbf{>~120\degree}                  & 7.77~(±1.48)   & 7.00~(±1.55)  & 2.46~(±0.47)    & 6.95~(±2.90)   & 7.72~(±2.90)   & 8.21~(±2.95)    & 6.50~(±1.54)   & 5.41~(±1.30) & 2.40~(±0.53)      \\
    \textbf{120\degree}                      & 9.28~(±2.20)   & 8.56~(±2.45) & 3.26~(±0.81)     & 8.92~(±3.64)   & 9.86~(±3.95)   & 10.35~(±4.00)   & 7.67~(±1.77)   & 6.49~(±1.54) &3.00~(±0.66)      \\
    \textbf{90\degree}                       & 10.16~(±2.13)  & 9.53~(±2.28)  & 3.88~(±0.91)     & 10.20~(±4.08)  & 10.90~(±4.10)  & 11.23~(±4.17)   & 8.12~(±1.93)   & 7.02~(±1.54) & 3.41~(±0.90)  \\
    \hline 
    \bottomrule

    \end{tabular}
    }
    
    \Description{The table lists the tracking errors of HOOV under three different horizontal headset FOV ranges. The caption of the table discusses the results in detail.}
    \label{tab:track_acc}
\end{table*}

\subsubsection{Success rate}

\autoref{fig:boxplot_error} (left) shows an overview of participants' performance success during this study.
Using \oculus, participants were careful not to make mistakes when performing the tasks under visual control.
Participants reached an average success rate of 94.12\% for \drop\ (max=100.00, min=88.24, SE=1.30), 94.61\% for \grab\ (max=100.00, min=79.41, SE=1.69), and 94.12\% for \compound\ (max=100.00, min=88.24, SE=0.93).
Using \optitrack, participants' mean success rate was 90.93\% for \drop\ (max=100.00, min=76.47, SE=1.91), 90.69\% for \grab\ (max=97.06, min=82.35, SE=1.50), and 90.56\% for \compound\ (max=100.00, min=83.82, SE=1.26).  

When simulating the tracking of participants' wrist motions outside the FOV with \projname{}, participants' overall success rate was 87.50\% for \drop\ (SE=1.63, max=97.06, min=76.47), 84.56\% for \grab\ (SE=2.02, max=97.06, min=67.65) and 83.82\% for the \compound\ task (SE=1.83, max=97.06, min=72.06).

\subsubsection{Trial Duration}

As shown in \autoref{fig:boxplot_time} (left), participants' mean completion time was shorter for the three tasks when not looking at the outside FOV regions, i.e., during \grab, \drop, and \compound\ in the \optitrack\ condition.
On average, participants took 2.37\,s to complete a \drop\ trial (max=3.27, min=1.66, SE=0.14), 2.65\,s for a \grab\ trial (max=3.50, min=1.96, SE=0.15), and 2.79\,s for a \compound\ trial (max=3.59, min=2.07, SE=0.17).
As shown in \autoref{fig:boxplot_time}, participants' mean completion time increased across all tasks in the \oculus\ condition.
Here, the average mean trial completion time was 2.93\,s for \drop\ (max=5.36, min=2.29, SE=0.22), 3.12\,s for \grab\ (max=4.58, min=2.33, SE=0.17), and 3.29\,s for the \compound\ task (max=4.36, min=2.57, SE=0.16).

\subsubsection{Absolute tracking accuracy of \projname}

\begin{figure}[b]
    \centering
    \includegraphics[width=\columnwidth]{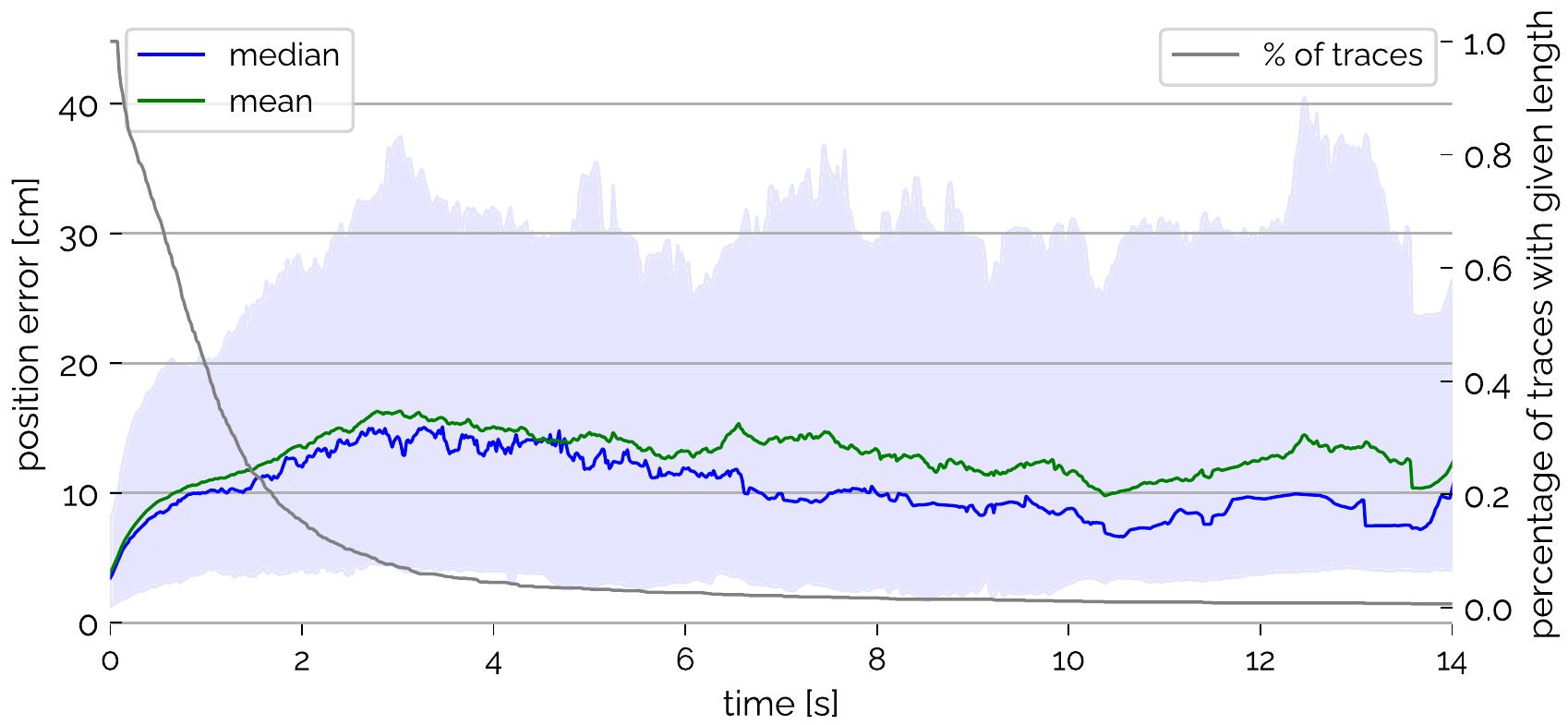}
    \vspace{-5mm}
    \caption{\hoov's tracking accuracy over time (compared to \optitrack\ tracking with sub-mm accuracy) in \studyone{}.
    Participants' interactions outside the FOV mostly occurred within the first three seconds, during which \projname\ achieved the best tracking performance and which covers the majority of interactions outside the user's field of view~\cite{wu2021user}.
    \hoov's tracking also remains stable afterwards, where the median and mean position error does not significantly exceed 15\,cm. The mean (green) and median (blue) are computed across the set of traces corresponding to a given duration of time outside the FOV. The area shaded in blue indicates the error interval between the 5th and 95th percentile.}
    \Description{The figure shows a graph depicting HOOV's tracking error over the duration of an out-of-FOV interaction. It also indicates the distribution over the out-of-FOV durations. Most of these interactions take less than three seconds. The mean and median tracking error both increase within this time span from around 3 cm to around 15 cm. They do not significantly exceed this error for a tracking duration of up to 14 seconds. The 95th percentile for the tracking error wanders around 30 cm.}
    \label{fig:time_error}
\end{figure}

Comparing the wrist poses estimated by \projname\ to the poses captured by the OptiTrack system when participants' hands were outside the tracking FOV of the Quest~2, we found an MAE of 7.77\,cm across participants.
The median position offset averaged across participants was 7.00\,cm.
\autoref{fig:time_error} illustrates the development of the average tracking error over time from the point when a participant's hand left the FOV of the headset.

When dropping an object outside the FOV, \projname's position estimates had an average MAE to the actual hand position of 6.95\,cm.
The average MAE when selecting objects outside the FOV was 7.72\,cm.
The orientation error as measured by the mean angle of the difference rotation is 6.50\degree\ (see~\autoref{tab:track_acc}).

As mentioned above, we simulated a series of tracking FOV to assess how \projname's tracking capabilities may deteriorate for larger areas outside the FOV and, thus, increased of \projname-based tracking.
For a simulated horizontal FOV of 120\degree, the average positional MAE across participants increased to 9.28\,cm, and the average MAD increased to 7.67\degree.
The offset to the actual release position and grab position while performing the \drop\ and \grab\ stayed within 10\,cm, amounting to 8.92\,cm and 9.86\,cm, respectively, on average.
For a simulated horizontal FOV of only 90\degree, \projname\ estimated positions with an average MAE of 10.16\,cm and estimated orientations with an MAD of 8.12\degree.

\subsection{Discussion}

The results of \studyone{} underline the potential of our method \projname\ to complement camera-based headset tracking for motions and interactions outside their tracking FOV.
Most obviously was participants' speed increase when they could directly interact with targets outside their FOV without turning their heads to first bring them into the view.
On average, mean completion times were 19\% lower for the \drop\ and 15\% lower for the \grab\ and \compound\ task when participants were able to interact outside their FOV.
Because participants were able to directly compare the conditions during their experience in the study, several mentioned, unprompted, that turning their heads felt like a burden during this task.

The results are also promising, as our study task simulated a common scenario in real life---performing an action for hand-object interaction while focusing on another object or action somewhere else.
Real-world examples include grabbing a water bottle or shifting gears while driving, or switching brushes while painting.

However, the gain in speed came at the cost of accuracy as expected.
Under visual control, participants achieved higher success rates for grabbing and dropping targets than operating outside their FOV without vision, which has been quantified in previous experiments~\cite{yan2018eyesfree, wu2021user, zhou2020eyes}.
Note that our targets were world-anchored and static; while the experimenter ensured that participants did not move from the position on the floor, shifts in upper-body pose may already result in a considerable effect.

Upon closer inspection of where the outside FOV input events were least correct, we saw a disproportional amount of errors for the spheres and drop zones behind the user.
In these cases, participants found it difficult to locate their hand outside their field of view correctly, possibly because these locations were outside participants' proprioceptive range and thus the area they would naturally interact with outside their FOV in other use cases.
Related, in erroneous cases, \projname\ failed to detect the correct object especially when participants placed their hand just between two targets.
Then, even small deviations in estimated positions led to binary changes of target outcomes.
We note that future systems may take this into account and potentially increase the separation between targets to correct for the tracking inaccuracies, introducing a dead band between targets where input events have no effect.

In terms of \projname{}'s potential to complement the tracking capabilities of headsets with limited FOVs, we first highlight its small median error of <~8\degree\ for all three simulated FOV conditions in~\autoref{tab:track_acc}.
Several applications in VR benefit from stable orientation estimation, such as games (e.g., Beat Saber) or 3D interactive environments (e.g., for ray-casting).
Next, we underline \projname{}'s capability to estimate \emph{absolute positions} outside the headset's tracking FOV.
Given a suitable interface design for virtual content outside the user's FOV, the median error of 7\,cm---about the length of an index finger---offers sufficient tracking accuracy to navigate through a range of targets in the hemisphere defined by the arm's length.
For this purpose, elements would need sufficient size and spacing that would allow users to traverse them in conjunction with the haptic feedback rendered by our wrist device.

While \projname{}'s tracking estimates are closest to the ground-truth within the first three seconds after the hand leaves the headset's FOV, \projname{} showed promise in that its tracking error deteriorated only minimally under our simulated 90\degree\ tracking FOV condition.
The mean position error of $\sim$10\,cm and the mean angle of the difference rotation of $\sim$8\degree offer encouraging benefits for \projname{} to work in conjunction with future headsets optimizing on power and form factor.

\begin{figure}[b]
    \centering
    \includegraphics[width=\columnwidth]{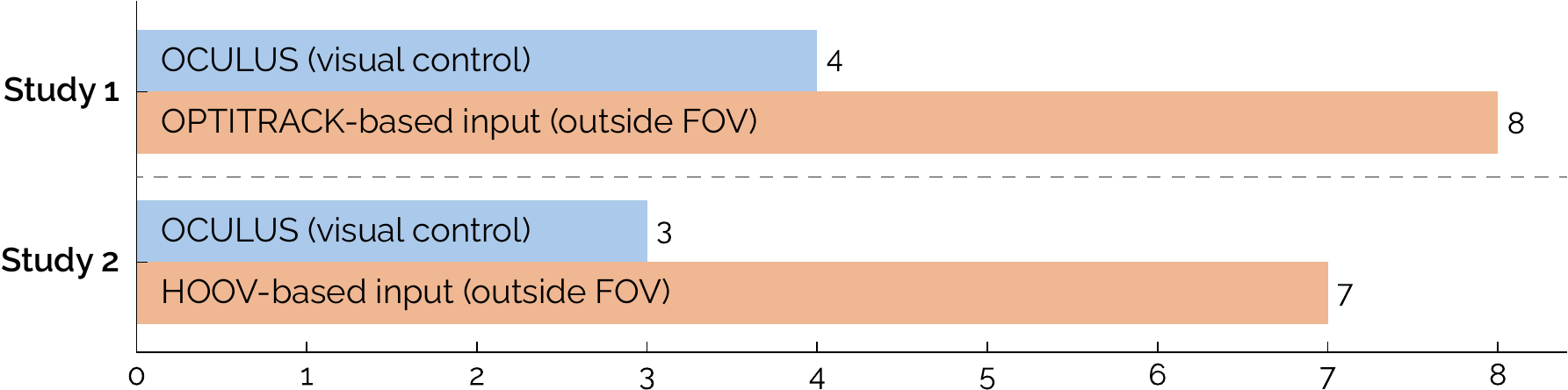}
    \vspace{-4mm}
    \caption{Participants' preferences of using one input technique over another.
    (Top)~In \studyone{}, more participants preferred interacting in the  \optitrack\ condition than vice versa, as it did not require them to turn their heads.
    (Bottom)~In \studytwo{}, this trend remained, even though interaction outside the FOV was now driven by \hoov\ in real-time.}
    \Description{The figure shows two bar graphs comparing participants' preferences of using one input technique over another in Study 1 and Study 2. In Study 1, 4 participants preferred the Oculus condition (input under visual control), while 8 participants favored the OptiTrack-based input (outside FOV input). For Study 2, 3 participants preferred interacting in the Oculus condition, while 7 participants would rather use the (outside FOV) HOOV-based input.}
    \label{fig:survey}
\end{figure}

Regarding participants' qualitative feedback (\autoref{fig:survey}), 
their overall reception of performing the tasks in the \optitrack\ condition was positive.
8 out of 12 participants reported that they preferred the \optitrack\ condition over \oculus, as their necks felt strained from repetitively turning around.
Several participants also mentioned that this condition felt more efficient, even if we saw during the analysis that they tended to make more mistakes.
In contrast, 4 of the 12 participants said that they would rather look at the object while performing the task, because it felt `safer.'

\paragraph{Comparison to related approaches}

\begin{table}[t]
    \caption{The table lists the median error (\emph{median pos.}) and mean absolute error (\emph{mean. pos}) of the tracked wrist position for \projname{} and other approaches reported in the literature in cm. It also shows the sensors used for tracking the wrist position. Note that the reported values from the literature rely on a Kinect 2.0 as the ground-truth reference sensor. Compared to related approaches that evaluate their tracking performance on recorded data \emph{offline}, we also analyze \projname{} in an \emph{interactive} study where participants complete tasks based on \projname{}'s tracking estimates.}
    \resizebox{\columnwidth}{!}{
    \centering
\begin{tabular}{lllcc} \toprule
 \textbf{Approach} & \textbf{Sensors} & \textbf{Evaluation} & \multicolumn{1}{l}{\textbf{median pos.}} & \multicolumn{1}{l}{\textbf{mean pos.}} \\ \hline
 ArmTrak (offline)~\cite{shen2016smartwatch} & 9-axis IMU & offline &  9.2 & - \\
 ArmTrak  (real-time)~\cite{shen2016smartwatch} & 9-axis IMU & offline &  13.3 & - \\
 MUSE~\cite{shen2018closing} & 9-axis IMU & offline &  8.8 & - \\
 MUSE (MR: grab-reach)~\cite{shen2018closing}  & 9-axis IMU & offline &  $\sim$15 & - \\
 LimbMotion~\cite{zhou2019limbmotion}  & 9-axis IMU (+edge detector) & offline &  8.9 & - \\
 RNN~\cite{wei2021real} & 6-axis IMU &  offline & 15.4 & 16.7 \\
 \projname{}---\studyone{} (Ours)& 6-axis IMU (+initial trajectory) & offline & \textbf{7.77} & \textbf{7.00} \\
 \projname{}---\studytwo{} (Ours)& 6-axis IMU (+initial trajectory) & \textbf{interactive} & 15.11 & 16.97 \\
 \hline 
\bottomrule

\end{tabular}
    }
    
    \Description{The table lists the median and mean tracking error of HOOV in comparison to related approaches.}
    \label{tab:track_comp}
\end{table}

\projname\ is not directly comparable to other methods that estimate the wrist position from a single IMU, because we evaluated our system on a different dataset and \projname\ receives an initial position of the hand as input.
However, \projname's relatively small median position error of 7\,cm demonstrates the strength of our estimation pipeline to handle the accurate estimation of the wrist pose over short periods.
Comparing the results to the values reported in the literature, \projname\ performs 15.5\% better than ArmTrak's offline algorithm~\cite{shen2016smartwatch}, 11.7\% better than the Particle Filter-based approach by \citeauthor{shen2018closing}~\cite{shen2018closing}, 12.7\% better than LimbMotion~\cite{zhou2019limbmotion}, which uses an additional remote edge device, and 49.5\% better than the RNN proposed by~\citeauthor{wei2021real}~\cite{wei2021real} (see~\autoref{tab:track_comp}).

In terms of human performance, the results from our study generally align with the conclusions drawn by~\citeauthor{yan2018eyesfree}~\cite{yan2018eyesfree}.
However, we also showed that participants were able to differentiate between objects in the lateral direction outside their field of view.

In order to understand \projname's estimation performance during real-time use, we conducted another evaluation where participants' interaction was based solely on \projname's tracking.

\section{\studytwo{}: Evaluation where participants interacted based on \projname's real-time estimates}
\label{sec:online_evaluation}

In \studytwo{}, we conducted another evaluation where participants interacted based on \hoov's real-time predictions, such that any potential tracking error of our method affected their behavior and thus performance.

\subsection{Study design}

\subsubsection{Apparatus}

We built on the experiment apparatus from our previous study (Section~\ref{sec:studydesign}).
To additionally evaluate the accuracy of our pinch detection and position, we attached two motion capture markers to participants' thumb and index fingernails before the study.
Tracking them, we simultaneously assessed the detection accuracy of pinch gestures as well as the accurate position of the fingers relative to the tracked wristband.
Our apparatus detected pinch events when the two markers converged to a distance below 3\,cm in a sudden motion.

In contrast to \studyone{}, the OptiTrack system was merely present in our apparatus to record ground-truth wrist poses for later analysis and had no influence on participants' behavior, the processed input events, or the virtual scene during the study.
Likewise, the apparatus continued to detect pinch events based on the IMU signals (and not based on the attached OptiTrack markers).

The VR front-end of our study apparatus received the wrist position and orientation from \projname's real-time estimations.
For this, a PC (32\,GB RAM, 3.60\,GHz CPU) with an NVIDIA GeForce RTX 3090 received the headset's pose, hand positions from the VR headset when visible, and the stream of IMU signals from \projname's wristband and processed them in real-time using the model trained on all data from \studyone{} (\autoref{sec:comparehoovopti}).

\subsubsection{Participants}

We recruited a fresh set of 10 participants (4 female, 6 male, ages 22--28, mean=25.3 years).
Headset distances from the ground were 157---190\,cm (mean=170\,cm, SD=9\,cm), and arm lengths ranged between 57---76\,cm (mean=69\,cm, SD=6\,cm).
Participants' prior experience with VR technology was lower than in our first study, ranging between 1--4 on a 5-point Likert scale (median=2).
Participants received a small gratuity for their time.

\subsubsection{Task} 

Participants completed the same three grab-and-place tasks (\drop, \grab\, \& \compound) as in \studyone{} while standing at an indicated stationary position.
The placement of the 17~virtual spheres and drop zones was also identical to our first study.

\subsubsection{Conditions}

Participants again completed each task in two different conditions.

The first condition was identical to the \oculus\ condition in \studyone{} (Section~\ref{sec:studydesign}), where participants visually followed their hands by turning their heads to interact with out-of-view targets.

In the second condition \hoov, participants faced straightforward and completed the tasks without turning their heads.
Unlike in \studyone{}, our apparatus tracked participants' interaction, rendered positions, and executed input events based on \hoov's predictions.
Participants again received haptic feedback at grab and release events and audio feedback depending on the outcome of a trial.

\subsubsection{Procedure}

At the start of the study, participants received an introduction to the tasks, provided information about their age and gender, and performed the T-pose calibration procedure to adjust the virtual environment to their body size.
Participants received training for each condition and task that included visual feedback for the \hoov\,condition.

Overall, participants completed 272\,trials (= 2 conditions $\times$ (34 \drop\ trials + 34~\grab\,trials + 68~\compound\,trials)) as fast as they could while avoiding errors.
Each drop zone and sphere appeared as target twice for each combination of task and condition.

\subsubsection{Measures}

Using our study apparatus, we measured the duration of each trial as well as participants' success rates.
We evaluated the tracking performance of \hoov\ compared to the ground-truth tracking from the OptiTrack's marker-based tracking system.
Using the additional optical markers, we also analyzed the accuracy of our pinch gesture detection.

\subsection{Results}

\subsubsection{Success rate}

Using the \oculus, participants correctly performed the task in 97.65\% (max=100.00, min=91.18, SE=0.83) of all trials for \drop, 95.88\% (max=100.00, min=85.29, SE=1.43) for \grab, and 97.94\% (max=100.00, min=92.65, SE=0.72) for \compound.

Using \hoov's real-time tracking, participants correctly completed the task in 87.65\% (max=97.06, min=70.59, SE=2.04) of trials for \drop, 86.18\% (max=100.00, min=73.53, SE=2.28) for \grab, and 86.18\% (max=91.18, min=76.47, SE=1.23) for \compound.

Analyzing the sub-mm accuracy measurements from the motion capture system for comparison, participants' hands were in the correct position 92.35\% (max=97.06, min=76.47, SE=1.83) of the trials for \drop, 95.59\% (max=100.00, min=79.41, SE=1.75) for \grab\ and 92.21\% (max=100.00, min=72.06, SE=2.09) during the \compound\ task.
Note that this analysis of human performance during interaction based on proprioception is hypothetical, as any real-time tracking error incurred by \hoov\ had an impact on participants' behavior during this study.

\subsubsection{Trial duration}

Using \oculus, participants on average completed trials in
3.17\,s (max=4.06, min=2.10, SE=0.20) for \drop,
3.74\,s (max=5.26, min=2.48, SE=0.22) for \grab,
and 3.90\,s (max=6.41, min=2.49, SE=0.32) for \compound.
Using \hoov's real-time tracking and without participants turning their heads, mean trial duration was
2.99\,s (max=3.63, min=2.44, SE=0.11) for \drop,
3.54\,s (max=4.07, min=2.93, SE=0.10) for \grab,
and 3.38\,s (max=4.11, min=2.87, SE=0.13) for \compound.

\subsubsection{Pinch accuracy}

Compared to the detected pinches using the two additional optical markers, \hoov's pinch detection achieved a recall of 94.85\% and a precision of 98.05\% across all three tasks.
Any false-negative detection in \hoov\ during the study led participants to pinch again, since \optitrack\ tracking had no impact on the behavior of the visual environment and study procedure.

\subsubsection{Absolute tracking accuracy of HOOV}

When considering \hoov{} as a continuous tracker of wrist position, \hoov's median position error was 15.11\,cm (MAE=16.97\,cm) averaged across all trials and participants.
In terms of accuracy during input events outside the FOV, the position error of \hoov's predicted release and grab events was
14.43 ($\pm$4.16)\,cm for \drop,
16.41 ($\pm$4.21)\,cm for \grab,
and 14.85 ($\pm$3.19)\,cm for \compound.

Analyzing \hoov's refined estimates of fingertip positions (15\,cm offset to the wrist) using the markers attached to participants' fingernails, 
the average error was 4.9\,cm when using the \optitrack's wrist position as the basis
and 14.0\,cm when using \hoov's estimated wrist orientation and position as the basis for refinement (which encompasses \hoov's error of predicting wrist positions).

\subsubsection{Qualitative feedback}

As shown in \autoref{fig:survey} (bottom), participants again preferred performing the study using \hoov\ overall, that is without requiring them to turn their heads to bring off-screen targets into view first.
7 out of the 10 participants reported a preference for the \hoov\ condition over \oculus.
Participants again mentioned the efficiency of not having to look for close-by off-screen targets before selecting them, although our analysis again showed that they had a slightly decreased success rate.

\subsection{Discussion}

Overall and compared to \studyone{}, participants were more careful in performing tasks in \studytwo{}, both in the \oculus\ and the \hoov\ condition.
This is evident as the success rates for \oculus\ and \optitrack\ both exceeded participants' results in the first study using the OptiTrack system (see~\autoref{fig:boxplot_error}).
This also explains the longer average trial completion time in \studytwo{} as shown in \autoref{fig:boxplot_time}.

Specific to the \hoov\ condition, participants achieved a similar performance in terms of success rate.
This demonstrates the capability of our learning-based method to generalize to and support interaction for users unseen during training without the need for calibration or fine-tuning.
It also shows our method's suitability for proprioceptive interaction when possibly imperfect tracking influences user behavior and thus alters their interaction.
Further supporting our method was participants' qualitative feedback, which highlighted that \projname{}'s tracking capabilities that extend and complement those of a regular headset were well-received.

In terms of tracking accuracy, \hoov\ produced a higher position error than in \studyone{}.
Again, this likely resulted from the longer time that participants spent completing trials and, thus, the increased amount of time they interacted outside the field of view.
This result is also commensurate with our assessment of tracking error over time as shown in \autoref{fig:time_error}.
Nevertheless, the absolute error in position remained relatively stable over time and still enabled participants to complete the intended tasks successfully.

The results from \studytwo{} also highlight the suitability of our pinch detection.
Our system is tuned towards a more conservative prediction of pinch events, resulting in a low number of wrongly detected peaks.
This rarely required participants to pinch more than once when our online detection missed an event.

Finally, this evaluation also validated our previous assumption of a constant offset between \hoov's predicted wrist position and the participant's fingertip.
The offset we found from the pinch position reported by the optical tracker was moderate and well below the error introduced by interacting based on proprioception outside the FOV in the first place.

\paragraph{Comparison to related approaches.} 
The wrist position error for the interactive tracking in \studytwo{} is comparable to the results reported for \citeauthor{wei2021real}'s RNN-based method~\cite{wei2021real}.
However, their results were obtained through an offline evaluation (similar to \studyone{}) using a Kinect camera with limited accuracy, and the displayed user motions had a limited range in the yaw direction.
While our performance falls short of the reported median tracking error of 8.8 cm for MUSE~\cite{shen2018closing} in a static setting with unsupervised user motions, the error becomes comparable in a medical rehabilitation (MR) application involving reach-and-grasp tasks (see~\autoref{tab:track_comp}). In addition, MUSE relies on a 9-axis IMU with a magnetometer prone to interference and performs the tracking in a 5-DoF space with respect to the torso.

\section{Applications}

Being complementary to the optical tracking abilities of commercial headsets, a multitude of applications can benefit from \projname\ during operation to support outside FOV interaction. 
A large number of game titles and immersive experiences rely on external tracking to obtain the position of the user's hand-held controllers around the body.
Such external tracking also supports comfortable interaction, where users can let their arms hang loosely by their sides, but still engage in an interactive experience thanks to controller input.
With \projname{}, we aim to enable similar levels of interactivity while supporting the portability-optimized form factor of current VR headsets, which therefore track all input inside-out as opposed to relying on additional sensing infrastructure.
To that end, we prototyped three demonstrations that illustrate the benefits of \projname{} in various application areas.

\subsection{Beat Saber}
To first illustrate the use of our system's extended tracking capabilities, we extended the \emph{Beat Saber} gameplay to include omnidirectional targets. 
Rather than presenting targets to users from a single direction, our game scene contains four tracks. 
Since users do not have a full visual understanding of the arriving targets, they have to rely more greatly on cues from the music to execute movements. 
For instance, they may be encouraged to swing backward on a regular beat. 
With inside-out tracking capabilities of HMDs, these movements would be difficult to track, precluding such interactions from being implemented in current applications. 
\projname{} resolves this issue and extends opportunities for gameplay to take advantage of a more dynamic range of movements.

\subsection{Archery}
Similar to Beat Saber, firing a bow requires movements that sometimes involve moving one's hand out of the tracking field-of-view of the head-mounted display. 
Unlike Beat Saber, archery further requires the user to steadily maintain an out-of-view hand position. 
\projname{} nonetheless supports this and further enables the release interaction with the on-wrist IMU. 

\subsection{Block Builder}
Lastly, we implemented a block builder application to demonstrate the applicability of \projname{} in areas beyond games and play. 
The Block Builder application consists of a workbench that is instrumented with containers for blocks around the edges.
Users can grab specific blocks from each container and piece them together on the workbench. 
As in a variety of other everyday activities, such as typing, we do not necessarily interact with the world with visual control, but rather sometimes rely on our sense of proprioception for control.
As users gradually gain familiarity with the environment, we can expect users to rely more so on their spatial memory to grab items as opposed to visual control. 
\projname{} enables this interaction.
\section{Limitations and Future Work}

While the results of the method we introduced in this paper are promising, several limitations exist that deserve further investigation in future work.

\paragraph{\projname\ as a method to compensate for IMU drift}
Combating the drift arising from estimating positions based on observations from inertial motions is a challenging problem.
With \projname, we so far address this for the limited space of a person's reach, only for a short period of time, and only for stationary and standing scenarios.
\autoref{fig:time_error} shows the development of tracking error over time, highlighting the challenge and remaining work needed to improve dead reckoning based on IMU signals.
\projname\ can therefore not be considered a general-purpose hand position estimator using inertial sensing.
The results we presented in this paper also assume reaching and grabbing tasks and may therefore not translate to other tasks users may perform in their proprioceptive reach, such as gesturing, drawing, or other kinds of spatial navigation.
The assumptions about the task, space, and scenarios we made to develop \projname\ specifically for stationary VR scenarios considerably set apart the problem we addressed from more general inertial-based odometry, which attempts dead reckoning over much larger areas in mobile and moving settings.

\paragraph{Non-stationary environments}
During our evaluation, participants stood still at a fixed position in the experimental space while performing all tasks as instructed.
Though calibrated to each participant's body dimensions once at the beginning of the experiment, all targets were static and remained world-anchored.
Although the experimenter verified throughout that participants did not step away from this position, the nature of the task did not allow us to control for a completely steady posture.
Therefore, shifts in upper-body posture or slight shoulder rotations may have contributed to the errors we observed, as participants obtained no feedback about their position in the virtual space relative to the targets.

Future evaluations could extend our evaluation to include non-stationary environments where participants can move around.
This would additionally evaluate the robustness of the method to motion artifacts.
Because \projname's algorithm currently needs a PC with a good GPU, such an evaluation would require a mobile reimplementation of our method, using model-size reduction techniques and deployment on the embedded platform to support such operation.

\paragraph{Computational complexity}
Due to the self-attention layers, \projname{}'s neural network incurs a complexity that grows quadratically with the length of the input sequence, limiting the maximum outside-the-field-of-view tracking duration.
Future work should consider architectures with recurrent elements and memory~\cite{bulatov2022recurrent} to avoid the repeated processing of the whole input sequence.

\paragraph{Tracking accuracy}
While the tracking accuracy of \projname\ is state-of-the-art for the given input modality, running from IMU-based orientation estimates only given the last hand observation from the headset, the existing error leaves room for improvement, especially compared to the tracking capabilities of an optical tracking system or a head-mounted display.
We believe that a part of this challenge can be addressed by acquiring a much larger data set, which would allow our method to be refined for personalization and to adapt to varying body sizes.

\paragraph{Instrumentation of the wrist}
\projname\ requires the user to wear a wristband with an embedded IMU to be tracked outside the FOV and to submit input commands.
While this is a common sensor inside every smartwatch today, it still requires a separate device to operate (parts of) VR experiences, which is counter to the intuition of manufacturers to integrate all VR components into just the headset.
Given the existing ecosystem of watches and their various programming interfaces, we are excited about the possibility of transferring and adapting our method to run on commodity devices and to support out-of-the-box use.
This will necessarily entail optimization strategies on the performance level as mentioned above in order to reduce the computational cost of our model or outsource its estimations to the neural computing unit of a headset.

\begin{figure}[b]
    \centering
    \includegraphics[width=\columnwidth]{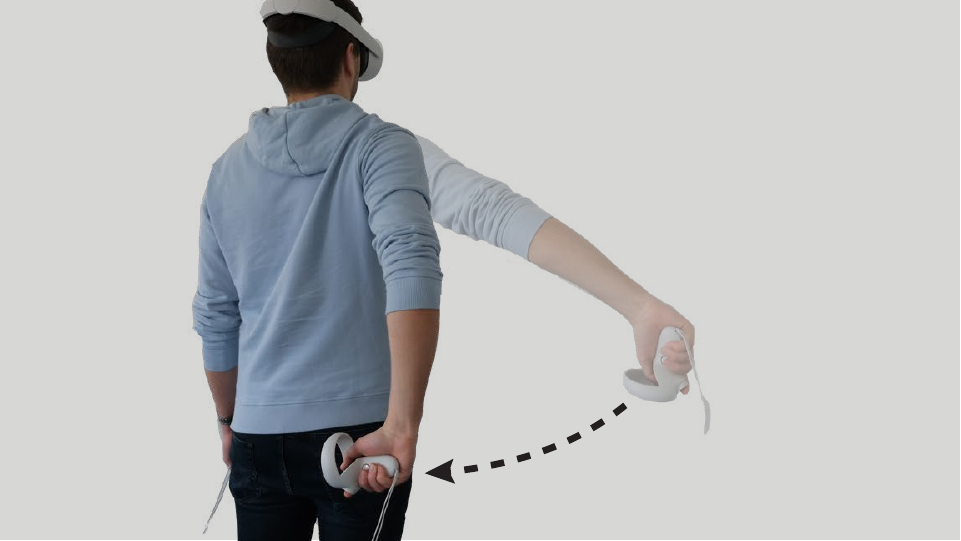}
    \caption{\hoov's wrist band contains the same sensor that is commonly found in hand-held controllers, allowing our method to generalize to a variety of input controllers.}
    \Description{The figure shows a user wearing an Oculus Quest 2 headset. The user is moving his right arm behind his back while holding a VR controller in his hand.}
    \label{fig:hoov_controller}
\end{figure}

\section{Conclusion}

We have presented \projname, a wrist-worn sensing and position estimation method that complements existing AR and VR headsets in tracking the user's hand interactions outside the tracking field of view of their built-in cameras.
\projname\ obtains its input from a 6-axis IMU that is embedded in a wrist-worn band, which captures 3-axis acceleration and 3-axis rotational velocity.
From these observations, \projname's learning-based inference pipeline estimates the user's current wrist orientation and position as soon as the hand leaves the tracking range of the headset.

Our user studies showed the promise of our method, allowing participants to leverage their proprioception and interact with one out of 17 targets outside their FOV with an overall mean success rate of around 85\% (\studyone{}) and 87\% (\studytwo{}).
The outside FOV interaction in our study supported speed improvements over what would be required on today's standalone headsets---turning one's head to bring the virtual content into view first and exclusively interact with it \emph{under visual control}.
While under visual control, participants successfully interacted with the correct targets in 19 of 20 cases, we quantified the drop in success rates under outside FOV operation:
Monitored with an external motion capture system, participants' success rate dropped to 18 of 20 trials when interacting purely based on their proprioception (and with targets that remained static and world-anchored).
Of these 18 successful trials, $\sim$17~interactions were correctly detected using \projname---without any external infrastructure or cameras.

Our study also uncovered \projname{}'s potential as a tracking complement for raw 3D positions outside the headset's tracking FOV.
\projname's median tracking error of 7\,cm for short-term outside-field-of-view interactions outperforms related techniques while requiring a commodity sensor that is commonly found in smartwatches and current hand-held controllers (\autoref{fig:hoov_controller}).

\projname\ opens up an exciting opportunity to broaden the interactive space for AR and VR headsets that may now be able to better leverage users' proprioception within the large outside FOV space around the user that can accommodate convenient reach.

Collectively, we believe that \projname\ will support the development of a future generation of AR and VR headsets that optimize for form factor and power consumption by outsourcing some of the computation to other devices.
We see an opportunity for adaptive interaction techniques---as commonly used for selection tasks where input is ambiguous (e.g., touchscreen typing)---to operate in conjunction with \projname{}'s estimates to support users' proprioceptive interactions.
Even more, we see the future potential for \projname\ to support hand tracking and interaction \emph{inside} the headset's FOV, thereby alleviating the computational cost of performing computer vision-based camera processing in real-time with a hand position estimation that may run on an embedded wrist-worn device.

\begin{acks}
We sincerely thank Manuel Meier for helpful discussions and comments.
We are grateful to NVIDIA for the provision of computing resources through the NVIDIA Academic Grant.
We thank the anonymous reviewers and all participants of our user studies.

\end{acks}

\balance
\bibliographystyle{ACM-Reference-Format}
\bibliography{hoov}


\end{document}